\documentclass[useAMS,usenatbib]{mn2e}
\usepackage{graphicx}
\title[Population synthesis of double neutron stars. ]{Population synthesis of double neutron stars.}
\author[S. Os{\l}owski et al.]{S. Os{\l}owski$^{1,2}$\thanks{E-mail:
soslowski@astro.swin.edu.au}and T. Bulik$^{3}$$^{,}$$^4$ and D. Gondek-Rosi\'{n}ska$^{4,5}$$^{,}$$^6$ and K. Belczy\'{n}ski$^3$$^{,}$$^7$\\
$^{1}$Swinburne University of Technology, Centre for Astrophysics and Supercomputing, Mail H39, PO Box 218, VIC 3122, Australia\\
$^{2}$CSIRO Astronomy and Space Sciences, Australia Telescope National Facility, P.O. Box 76, Epping, NSW 1710, Australia\\
$^{3}$Astronomical Observatory, University of Warsaw, Aleje Ujazdowskie 4, 00-478, Warsaw, Poland\\
$^{4}$Nicolaus Copernicus Astronomical Centre, Bartycka 18, 00716 Warszawa, Poland\\
$^{5}$Institute of Astronomy, University of Zielona G\`{o}ra, Lubuska 2, 65-265 Zielona G\`{o}ra, Poland\\
$^{6}$LUTH, Observatoire de Paris, Universite Paris 7, Place Jules Janssen, 92195 Meudon Cedex, France\\
$^{7}$Dept. of Physics and Astronomy, University of Texas, Brownsville, TX 78520, USA \\}
\begin{document}

\date{Accepted . Received ; in original form}

\pagerange{\pageref{firstpage}--\pageref{lastpage}} 
\pubyear{2010}

\maketitle

\label{firstpage}

\begin{abstract}

Using the S\textsc{tar}T\textsc{rack} binary population synthesis code
we model the population of double neutron stars in the Galaxy.  We include a
detailed treatment of the spin evolution of each pulsar due to processes such as
spin-down and spin-up during accretion events as well as magnetic field decay.
We also model the spatial distribution of double neutron stars by including
their natal kicks and subsequent propagation in the Galactic gravitational
potential.  This synthetic pulsar population is compared to the observed sample
of double neutron stars taking into account the selection effects of detection
in the radio band, to determine the most likely evolutionary parameters.  With
these parameters we determine the  properties of the double neutron star
binaries detectable in gravitational waves by the high frequency interferometers
LIGO and VIRGO.  In particular, we discuss the distributions  of chirp masses
and mass ratios in samples selected by their radio or gravitational wave
emission.

\end{abstract}


\section{Introduction}

Double neutron stars (DNSs) are remarkable objects in astrophysics and their
studies uniquely contribute to a number of areas of fundamental significance.
The monitoring of PSR 1913+16 provided strong limits on the General Theory of
Relativity \citep{1975ApJ...195L..51H}. The recent discovery of the first system
with two observable pulsars J0737-3039 (\citet{2003Natur.426..531B} and
\citet{2004Sci...303.1153L}) has been a breakthrough in pulsar studies. The
eclipses in this system provide insight in the physics of the magnetosphere
\citep{2005ApJ...634.1223L}.  Measuring the moment of inertia of pulsar A in
this system would lead to a very strong constraint on the neutron star equation
of state.  By studying the properties of DNSs, we
infer that a population of merging binaries exists. Other compact
object binaries do not provide evidence of sources of gravitational
waves potentially detectable by the LIGO and VIRGO observatories
(\citet{1992Sci...256..325A} and \citet{1991pfmp.conf..341B}).  The origin of
DNSs was first conjectured by \citet{1975A&A....39...61F} in the context of the
the Hulse Taylor pulsar PSR 1913+16. The birth rate and properties of DNSs were
investigated by \citet{1992AIPC..272.1626P}, and by \citet{1999ApJ...520..696A}.
\citet{2001ApJ...550L.183B} pointed out that there may exist a large population
of ultra-compact DNS binaries formed with an additional common envelope stage,
and not detectable in the radio.  The existence of this population relies
strongly on the result of the common envelope phase with a star on the
Hertzsprung gap.  For a review on binary pulsars see
\citet{2008LRR....11....8L}.

Pulsar evolution was first investigated by \citet{1970ApJ...160..979G} soon
after the discovery of pulsars. Since then, most papers concentrated on
modelling the observed sample of pulsars by including all selection effects and
modelling pulsar evolutionary parameters such as initial distributions, magnetic
field decay, evolution of inclination, etc. (\citet{1971IAUS...46..165L},
\citet{1977ApJ...215..885T}, \citet{1985MNRAS.213..613L},
\citet{1987A&A...178..143S}, \citet{1989ApJ...345..931E},
\citet{1992A&A...254..198B}, \citet{1997A&A...322..477H},
\citet{2001ApJ...556..340K} \citet{2002ApJ...568..289A},
\citet{2002ApJ...565..482G}, \citet{2004ApJ...604..775G},
\citet{2006ApJ...643..332F} ). These efforts concentrate primarily on classical
pulsars; more recently \citet{2007ApJ...671..713S} present models of the
population of recycled millisecond pulsars. \citet{2008MNRAS.388..393K,
2009MNRAS.395.2326K} are the first to  take into account the effects of binary
evolution leading to formation of these objects but assume initial injected
distributions of magnetic fields and spin periods.  Our work focuses on DNSs
only and is based on different stellar evolution codes. We address the issue of
pulsar populations detectable either in the radio or gravitational waves.

The evolution of binaries leading to formation of double compact objects has
been investigated by many authors \citep{1985SvA....29..645G,
1995MNRAS.274..461B, 1995Ap&SS.231..389J, 1995ApJ...454..593L,
1998A&A...332..173P, 2000A&AT...19..471P, 2005MNRAS.363L..71D,
2005ARep...49..295B, 2006MNRAS.368.1742D, 2008MNRAS.388..393K}.  Here we will be
using the Star Track binary population synthesis code which was presented and
used in a series of papers by Belczynski and collaborators (see
\citet{2002ApJ...572..407B}, \citet{2008ApJS..174..223B} and references
therein).  The code has been used to investigate the possible binary progenitors
of gamma ray bursts and to study the rates and properties of binary
gravitational wave sources \citep{2003ApJ...589L..37B,2004A&A...415..407B}.
\citet{2004MNRAS.352.1372B} and \citet{2007AdSpR..39..285G} presented a
calculation of the expected mass spectrum of merging compact object binaries
including double neutron stars. They hinted at a possibility of detecting a
large number of non equal mass DNS binaries, and showed that they may not show
up in the radio sample due to selection effects. 

In this paper we present the merger of binary population synthesis and pulsar
evolution models.  After applying radio selection effects we compare the
predictions with the available data. The main differences between our work and
that of \citet{2008MNRAS.388..393K} is that we base our modelling on different
stellar evolution codes, focus on DNSs only and choose different field decay
timescales.  In Section 2 we present our model of binary evolution and pulsar
evolution.  Section 3 contains the results and comparison with the observations.
Section 4 focuses on the mass distribution of NS visible in the radio and
gravitational waves, while Section 5 contains the conclusions.

\subsection{Properties of known double neutron star systems}

\begin{figure}
 \includegraphics[width=\columnwidth]{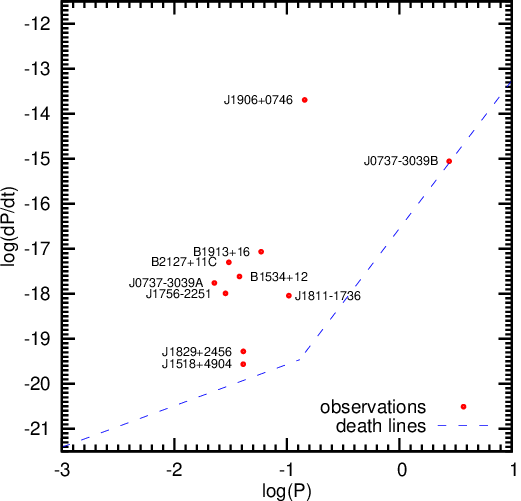}

\caption[Observed double neutron stars]{Observed double neutron stars on
$P-\dot{P}$ diagram.  Name of each pulsar is given next to it. We currently know
about 10 pulsars in DNS binaries.  In one case both are visible as radio pulsars
(J0737-3039). The death lines are described in Section \ref{radiolum}.}

\label{pulsobs}
\end{figure}

Currently there are ten known pulsars in nine DNS binaries. In one case, both
stars in the binary have been observed as radio pulsars (J0737-3039), but the B
pulsar is no longer visible due to the spin precession and expected to reappear
in the year 2035 \citep{2010ApJ...721.1193P}.  The spin period versus spin
period derivative diagram ($P-\dot{P}$) for these binaries is presented in Fig.
\ref{pulsobs}.  Most of the objects are concentrated in the region of the
partially recycled pulsars.  However, there are two noticeable outliers:
J0737-3039B, the companion of the J0737-3039A; and J1906-0746, a young pulsar
likely to have a neutron star companion.  The pulsar B2127+11C probably had a
very different dynamical history. It lies within the globular cluster M15, where
three body interactions could have replaced its original companion with a
neutron star and have kicked the pulsar further away from the cluster's core
\citep{1991Natur.349..220P}.  Detailed properties of these systems can be found
in Table \ref{pulsobswlas}, sorted by the time remaining to coalescence (merger
times). Three binaries at the bottom of the table have merger times much longer
than $10\;{\rm Gyr}$.  We list the spin periods and their derivatives, the
masses of the neutron star, as well as the present orbital parameters.  The
distribution of the orbital periods spreads over two orders of magnitude, with
no clear evidence of clustering in this range. The orbits  of all of the systems
are significantly eccentric. The masses of the seven objects with merger times
shorter than 10 Gyrs are very well determined and are in the range between 1.25
and 1.44~M$_{\sun}$. The masses of the neutron star in  the remaining three
binaries are not so well constrained and may even lie outside this range.  
 
\begin{table*}
 \begin{minipage}{165mm}

 \caption{Properties of the observed pulsars in DNS binaries. The table contains
names, spin period, spin period derivative, orbital period, mass of observed
neutron star, mass of the companion, eccentricity of the orbit and time to
merger. All given digits are significant. Errors, where given, are $1-\sigma$
errors. References: 1 - \citet{2004Sci...304..547S}, 2 -
\citet{2006ApJ...644L.113J}, 3 - \citet{2008AIPC..983..485K}, 4 -
\citet{2005ASPC..328...25W}, 5 - \citet{2005ApJ...618L.119F}, 6 -
\citet{Janssen}.}

 \label{pulsobswlas}
\begin{tabular}{|c|c|c|c|c|c|c|c|c|c|c|}
\hline
Name & $P\;[ms]$ & $\dot{P}\; [ss^{-1}/10^{-18}]$ & $P_{orb}$ [h]& $M_{obs}\;[M_{\sun}]$ & $M_{cmp}\;[M_{\sun}]$ & $e$ & $\tau_{mrg}\;[Gyr]$ & Reference \\
\hline
J0737-3039A& 22.70   & 1.74          & 2.454   & $1.337^{+0.005}_{-0.005}$ &  $1.250^{+0.005}_{-0.005}$ & 0.088 & 0.085 & 1 \\
J0737-3039B& 2773    & $8.8\times 10^2$ & 2.454 &  $1.250^{+0.005}_{-0.005}$ &  $1.337^{+0.005}_{-0.005}$ & 0.088 & 0.085 & 1 \\
B2127+11C  & 30.53   & 4.99          & 8.05    &  $1.358^{+0.01}_{-0.1}$ &  $1.34^{+0.01}_{-0.01}$ & 0.681 & 0.2 & 2 \\
J1906+0746	 & 144.07  & $2.028\times 10^4$& 3.98    & $1.248^{+0.018}_{-0.018}$ & $1.365^{+0.018}_{-0.018}$ & 0.085 & 0.3 & 3\\
B1913+16   & 59.03   & 8.63          & 7.752   & $1.4414^{+0.0002}_{-0.0002}$ &  $1.3867^{+0.0002}_{-0.0002}$ & 0.617 & 0.3 & 4\\
J1756-2251 & 28.46   & 1.02          & 7.67    & $1.312^{+0.017}_{-0.017}$ & $1.258^{+0.018}_{-0.017}$ & 0.181 & 1.7 & 5 \\
B1534+12   & 37.90   & 2.43          & 10.098  & $1.3332^{+0.001}_{-0.001}$ & $1.3452^{+0.001}_{-0.001}$ & 0.274 & 2.7 & 1 \\
J1811-1736 & 104.182 & 0.91         & 451.20  & $1.62^{+0.22}_{-0.55}$ & $1.11^{+0.53}_{-0.15}$ & 0.828 & $>$10 & 1\\
J1518+4904 & 40.935  & 0.027          & 207.216 & $0.72^{+0.51}_{-0.58}$ & $2.00^{+0.58}_{-0.51}$  & 0.249 & $>$10 & 6\\
J1829+2456 & 41.0098 & 0.05   & 28.0    & $1.14^{+0.28}_{-0.48}$ & $1.36^{+0.50}_{-0.17}$ & 0.139 & $>$10 & 1 \\
\hline
\end{tabular}
\flushleft

\end{minipage}
\end{table*}

\subsection{Radio pulsars and the population observable in gravitational waves}
\label{radgw}

Due to selection effects the observed sample of DNS binaries might differ from
the intrinsic population.  For the same reason the population observed in the
radio might have different properties than the one we would observe in the
gravitational waves. This has implications for the interpretation of the results
from laser interferometer gravitational waves observatories.

The most obvious selection effect in the radio is the inverse square law.  Our
radio-telescopes have a limited sensitivity, so we will not see the weakest
and/or farthest pulsars.  The DNS population observed in the radio is so far
restricted to our Galaxy.  In this paper we model the radio selection effects
due to the interstellar medium and the telescope properties (described in detail
in Section \ref{radiodet}).  In our calculations we incorporate another
selection effect which is due to the fact that the pulsars are in binaries.
When searching for pulsars with stack search and phase modulation methods, it is
difficult to detect radio emission from neutron stars in binaries with an
orbital period shorter than 4h \citep{Faulkner}. In general, this is because the
signal-to-noise ratio drops drastically due to variation of the pulsar period
during the observation. This orbital period limit was calculated for a pulsar
with a spin period $P=9\;{\rm ms}$ \citep{Faulkner}, while the limit may vary as
a function of $P.$ We took this effect into account in this approximate form,
but it has a very small impact on the results. Another important consequence of
this effect is the reduced number of discovered binaries in pulsar surveys.  In
addition, there are  different methods of detecting radio pulsars which are more
sensitive to pulsars of such properties. The so called acceleration searches
recover the signal-to-noise ratio by modelling the drift of the pulsar spin
period due to its orbital motion.  These methods are not always implemented in
searches for new pulsars in survey data, but progress in this area is being made
(e.g. \citet{2003ApJ...589..911R}).  In practice, the acceleration searches are
often not included in the standard pipeline processing of survey data since they
are very computationally expensive.

Our first aim is to investigate how the described selection effects influence
the observed properties of the radio pulsar population. We model the evolution
of pulsars in binaries and compare the results of different models with
observations. Binaries detectable in the gravitational waves are those with a
relatively small orbital separation. We assume that only binaries with a
coalescence time shorter than $10^{10}$ years are representative of the
population detectable in the gravitational waves. The strength of the
gravitational wave signal during the inspiral phase depends on the chirp mass of
the binary. This is the main selection effect for DNS's observability in the
gravitational waves (see Section \ref{gwdet}).  In \citet{2004MNRAS.352.1372B}
and \citet{2005MmSAI..76..513G} the authors suggest that there will be
differences between the DNS population observable in the radio and gravitational
waves. In particular, they conclude that the mass ratio distributions will not
be the same for the two populations.  The population modelling in those papers
was based on a number of invalid assumptions.  The binaries with lifetimes above
the Hubble time were neglected, and it was assumed that observability of pulsars
is proportional to their lifetime.  The gravitational wave strains are very low
and the signals that interferometric detectors are trying to detect are buried
in noise.  Numerical relativity calculations provide an input for statistical
methods that enable the detection of a known signal from the noise. Until
recently most calculations of coalescing neutron star and/or black hole binaries
assumed that the mass ratio is close to unity (e.g. \citet{2007AdSpR..39..271G},
\citet{2007PhRvD..75l4018B}).  In this paper we use the results of modelling the
radio selected population to obtain the mass ratio distributions of DNS systems.
We check whether the pulsars potentially detectable in the gravitational waves
have the same properties as the radio population.

\section{Description of the Model}

\subsection{S{\sevensize TAR}T{\sevensize RACK} binary population synthesis code}

In order to perform a population synthesis of double neutron stars we need to
know what the population of DNS binaries itself look like. We used the
S\textsc{tar}T\textsc{rack} code for this purpose \citep{2008ApJS..174..223B}. This code has
been improved for many years now and is considered to be the state-of-the-art
with regard to our current understanding of stellar evolution.
S\textsc{tar}T\textsc{rack} has
been extensively tested, calibrated and its results have been compared to
observations
\citep[e.g.][]{2002ApJ...571..394B,2002ApJ...574L.147B,2002ApJ...567L..63B,2008ApJS..174..223B}. At the base of this code are analytic formulae describing the evolution of single stars first
developed by \citet{Hurley:2000pk}. The code follows a sequence of evolutionary
phases of stars starting at the main sequence, passing through the Hertzsprung
gap, the red giant branch or the asymptotic giant branch. After its nuclear
evolution the star can form a compact object (a white dwarf, a neutron star or a
black hole). The supernova explosion can completely disrupt the star leaving no
remnant at all. S\textsc{tar}T\textsc{rack} includes the evolution of helium stars as well. If
the star that evolves first ends up as a white dwarf or a neutron star, then
there is a chance a double neutron star system will be formed, provided that the
binary is not disrupted during the supernova explosions.  In the S\textsc{tar}T\textsc{rack}'s
standard model we assume a bimodal distribution of the FeNi core masses in
pre-supernova stars (Weaver \& \citet{1993PhR...227...65W},
\citet{1996ApJ...457..834T}). Stars with initial masses below $\sim 18-19$
$M_{\sun}$ burn carbon convectively. They form cores of $\sim 1.5\;{\rm
M_{\sun}}$  which yield neutron stars of $1.3\;{\rm M_{\sun}}$.  Heavier stars
burn carbon radiatively and the cores are more massive $\sim 2\;{\rm M_{\sun}}$
(yielding neutron stars of $1.8\; {\rm M_{\sun}}$). The mass of the FeNi core
does not depend sensitively on the initial star mass, but is a strong function
of the burning reactions within the pre-supernova star. If similar reactions are
encountered over the wide initial star mass range in which neutron stars are
formed (e.g. 8-18 $M_{\sun}$), then the majority of neutron stars are expected
to have similar masses. The previously employed neutron star mass calculation
was based on stellar evolutionary models that suggest an increase of final FeNi
core mass with initial star mass (e.g.. \citet{1993PhR...227...65W}; or see
Table 4.4 of \citet{1998A&ARv...9...63V}). This naturally results in a wide
neutron star mass spectrum.

Some poorly known details of the stellar evolution are parametrised in the
S\textsc{tar}T\textsc{rack} code. Fortunately most of them do not have a significant impact on
the results relevant to this work. The various models of the binary evolution in
this code are described in Section~\ref{evolotbin}.

\subsection{Evolution of the binaries}
\label{evolotbin}

We consider three variants of the binary population synthesis models.  An
outline for each one of them is given below.

\subsubsection{Model A - standard}
\label{astand}

This model is described in detail in \citet{2002ApJ...572..407B} and updated in
\citet{2008ApJS..174..223B}.  Here we only present the most important features.
The initial mass of one of the stars is drawn from a power law $\sim M^{-2.7}$
\citep{1986FCPh...11....1S}. The mass of its companion is determined as a random
fraction of the heavier (at the time of formation) star's mass drawn from a flat
distribution \citep{1935PASP...47...15K}. The initial orbital separation
distribution is flat in logarithm, while the initial eccentricity's distribution
is flat.  The code calculates the carbon-oxygen core mass, which will form a
compact object using formulae from the already mentioned work by
\citet{Hurley:2000pk}. This mass has an impact on the baryonic mass of the newly
formed neutron star due to its influence on the mass of the Fe-Ni core as
described in \citet{1986nce..conf....1W}, \citet{1999ApJ...526..152F} and
\citet{1996ApJ...457..834T}
\begin{equation}
M_{NS}^{bar} = \left\{
\begin{array}{ll} M_{FeNi}, & M_{CO} \lid 5 M_{\sun} \\ M, & M_{CO} \gid 7.6
M_{\sun} \\ M_{FeNi}+f_b(M-M_{FeNi}), & {\rm otherwise} \\ \end{array} \right.,
\end{equation}
where $f_b$ is the fall-back factor (describing the fraction of the stellar
envelope that falls back after the SN), $M_{FeNi}$ is the Fe-Ni core mass,
$M_{CO}$ is the CO core mass and $M$ is the pre-SN mass of the star; for details
see \citet{2008ApJS..174..223B}.  The fall-back factor is a linear function of
$M_{CO}$, such that $f_b(M_{CO}=5M_{\sun})=0$, and $f_b(M_{CO}=7.6M_{\sun})=1$.
In order to convert the baryonic masses to the gravitational masses a quadratic
formula is used: $M_{NS}^{bar} - M_{NS} = 0.075M_{NS}^2$, where $M_{NS}$ is the
gravitational mass. Orbital parameters are affected by tidal interactions, mass
loss due to stellar winds, mass transfers, and gravitational wave emission. We
parametrise the fraction of the mass ($f_a=0.5$) taking part in a stable mass
transfer that is accreted by the companion; the rest is lost from the binary.
Unstable mass transfers in S\textsc{tar}T\textsc{rack} (common envelope - CE) is treated with
the formalism proposed by \citet{1984ApJ...277..355W} and
\citet{1990ApJ...358.189D}. In this model we assume $\alpha_{CE}\times\lambda =
1.0$. The neutron star can accrete some mass during the common envelope phase.
The amount of this mass is drawn from the range $[0.05 M_{\sun},0.10 M_{\sun}]$
with a flat distribution.  Each supernova explosion occurs at a random place in
the orbit. We take natal kicks drawn from the \citet{2005MNRAS.360..974H}
distribution into account, and calculate the orbit and centre of mass velocity
after the explosion.

\subsubsection{Model H}
\label{H}

Model H treats the common envelope phase  differently.  In this case neutron
stars accrete more matter by assuming the Bondi-Hoyle accretion
\citep{1944MNRAS.104..273B} with a hypercritical accretion rate
\citep{2000ApJ...541..918B}. It has a crucial impact both on the binary
evolution (changes in orbital separation) and on the properties of the pulsar
that formed first.  The amount of accreted mass is roughly $10$ times more than
in the standard model A. This amount depends mostly on the mass ratio in the
binary prior to the common envelope $q_0$. The final mass is calculated using
formula \citep{1998ApJ...506..780B}:
\begin{equation}
M_B=\left(q_0+0.7q_{0}^{2}\right)M_A\,,
\end{equation}
where $M_B$ is the mass of the acceptor after the transfer and $M_A$ is the mass
of the donor before the transfer.

\subsubsection{Model S}

In this model we assume a different initial mass function for the newly formed
neutron stars.  It assumes a very simplified linear relation between $M_{NS}$
and the mass of the iron core $M_{core}$ from which it is formed:
\begin{equation}
M_{NS}=0.35\times M_{core}+0.596M_{\sun}.
\end{equation}
The values of the parameters are such that we have neutron stars with masses
from $1.1M_{\sun}$ (forming from a star with mass $8.275M_{\sun}$ on a zero age
main sequence) to $2.5M_{\sun}$ (forming from a star with mass $20.88M_{\sun}$
on a zero age main sequence). The purpose of this model is to check the
influence of this poorly known relation on the distribution of the masses of the
double neutron stars.

\subsection{Pulsar evolution model}
\label{ewolucja}

The previous subsections describe the models of the stellar evolution and the
initial masses of the neutron stars. It remains to assign each neutron star
other properties such as magnetic field and spin period. These parameters are
required to model the neutron stars as radio pulsars.  The S\textsc{tar}T\textsc{rack} code
provides details of the interaction in the binaries. We assume that all neutron
stars become radio pulsars immediately after their birth. The pulsar that is
born earlier can be affected by accretion of matter from its still evolving
companion. These effects are modelled and described in the following paragraphs.

Before moving to the details of our model, we briefly describe the ingredients
necessary to model the life and death of each pulsar. At the very beginning we
assign each neutron star a set of initial parameters such as spin period,
magnetic field and moment of inertia. After that, we start evolving each pulsar.
We assume a dipole model for the magnetic field \citep{1969ApJ...157.1395O} and
base the spin-rate evolution on that.  Since we only consider double neutron
stars, we need to consider mass transfer and how it affects radio pulsars.
During accretion the neutron stars are visible in the X-ray and not detectable
in the radio. In this paper we do not model the X-ray emission as it only occurs
for a  short period of time compared to the radio activity periods. The
evolution of binaries, along with their movement in the Galaxy, is followed
until the coalescence due to gravitational wave emission or for $ 10\;{\rm
Gyr}$, whichever is shorter.

\subsubsection{Initial parameters}
\label{init}

Our model starts treating each neutron star as a radio pulsar immediately after
its formation. The initial magnetic field is drawn from a flat logarithmic
distribution in the range $10^{11}{\rm G} \lid B \lid 10^{13}{\rm G}$.

The initial spin period is, in most models,  $P_{ini}=10\; {\rm ms}$. The choice
of this parameter is not very important. Even if some (or all) of the pulsars
were born with longer or shorter spin periods, the overall properties of the
resulting population would not change significantly unless we vary the initial
period by more than a factor of 10. As a matter of fact, longer initial periods
will make it harder to get a  good fit to the observations. This will be
demonstrated by the model APD05I where we allow a uniform distribution of the
initial spin periods between 10 and 100 ms as motivated by some observations
\citep{2006ApJ...643..332F}. This will make the model slightly worse as it will
be harder to spin pulsars up to short enough spin periods.  Young pulsars spend
a very short time in the region of the $P-\dot{P}$ diagram where they are born
and move quickly towards longer spin periods.  Given the small number statistics
of the observed sample, it is not possible to distinguish the models with more
realistic $P_{ini}$ distributions. As a consequence and for the sake of
simplicity most of the models we present will have $P_{ini}=10\;{\rm ms}$ for
all the pulsars. We have tested models with longer initial periods, up to 100
ms, but we found 10 ms to be the optimal initial period.

The radius $R=10\;{\rm km}$ and moment of inertia $I=10^{45}\;{\rm g\times
cm^2}$ are the same for all the pulsars.

As mentioned before, we consider only spin-down due to dipole emission. This
yields the braking index $n=\ddot{\Omega}\Omega \dot{\Omega}^{-2}=3$ and
determines the spin frequency evolution. Observations show that the breaking
indices of pulsars are in the range $\approx 2.5-3.5$
\citep{2005yCat.7245....0M}.  Following the pulsar dipole model, the rotation of
the neutron star slows down with the initial frequency time derivative:
\begin{equation}
\dot{\Omega}=-\frac{2B^2R^6\sin^2\alpha}{3c^3I}\Omega^3\;,
\label{Omdot}
\end{equation}
where $\alpha$ is the angle between the rotation and magnetic field axis, $B$ is
the magnetic field, $R$ is the neutron star radius, $c$ is the speed of light
and $I$ is the moment of inertia of the neutron star. We assume $\alpha=30\degr$
for each pulsar, as values of $\alpha < 45 \degr$ have been shown to provide
better fits to observations \citep{2006ApJ...643.1139C}. The magnetic field
strength and the angle $\alpha$ are degenerate so we do not lose anything in
making this assumption. This degeneracy extends to the evolution of the field
and the inclination angle.  At every point in the evolution, the magnetic field
is given by
\begin{equation}
B=6.4\times 10^{19} \sqrt{P \dot P}\;{\rm Gauss}\, .
\end{equation}

\subsubsection{Evolution of the magnetic field}
\label{fielddec}

If there are no direct interactions between the binary members, the spin period
and its evolution are independent of the companion's properties. This is the
case after the formation of the first pulsar, unless there is ongoing mass
accretion, and after the formation of the second neutron star until the contact
of magnetospheres and coalescence. During these periods of independent
evolution, constituting most of the pulsar's lifetime, the spin period evolves
according to Equation ~\ref{Omdot}.  In addition we assume that the magnetic
field of a pulsar is decaying exponentially in time via Hall-like effect with
ohmic dissipation (\citet{1990Natur.347..741R,
1994MNRAS.271..490G,1995MNRAS.275.1117U,1997MNRAS.292..167U,
1997MNRAS.284..311K,1999MNRAS.303..588K,1999MNRAS.308..795K}, see also Section
\ref{roche} for other possiblities):
\begin{equation}
B=\left(B_0-B_{min}\right)\cdot\exp\left(-\frac{t}{\tau_d}\right)+B_{min}\;,
\end{equation}
where $\tau_d$ is the magnetic field decay timescale.  The decay of the
magnetic field is controversial. Some analyses show that it proceeds on a
timescale of $2-5\;{\rm Myrs}$ \citep{2002ApJ...565..482G,2004ApJ...604..775G},
while others \citep{2006ApJ...643..332F} argue that it may be an artefact of 
the assumed luminosity law. \citet{2008MNRAS.388..393K} obtains the best model
with a long ($\approx$ 2000 Myrs) timescale for the decay of the magnetic
field. In their model, the initial spin period is correlated with the initial
magnetic field. \citet{2010HEAD...11.1612G} have tested both possibilities and
finds that it is not possible to create a good synthetic population without
either quickly decaying the magnetic field or correlating the initial spin
periods  with the magnetic fields. This work suggests that it is easier to
reproduce the observed population in the first case and our models will follow
that path.  The short timescale of the magnetic field decay can be interpreted
as a way of modelling the deviations of the pulsar spin evolution from the
standard dipole model. This is supported by the fact that there seems to be
some additional torque in the pulsar evolution \citep{2006ApJ...643..332F}.
The caveat of this solution is that the second neutron star can pass through
the region of the partially recycled pulsars without accreting any mass.
However the fraction of the recycled pulsar's contribution to that region is
high and they dominate the results for most of the models (see Section
\ref{popnappdot}). While the main goal of this work is to look for 
differences between the populations visible in the radio and potentially 
detectable by gravitational wave detectors we also attempt to distinguishing
between different physical models of the magnetic field evolution and to create
an improved model of the pulsar spin-down.  We want our modelled pulsars to
evolve as described by the standard dipole model at later stages of their life;
therefore we assume that after the first born pulsar  accretes any amount of
mass its field ceases to decay. The models with very long timescale decays of 1
Gyr and more are an exception to that rule. In these cases the magnetic field
decays for the whole lifetime of the pulsar. In addition,
\citet{2006MNRAS.366..137Z} shows that there is a lower limit to the magnetic
field strength. We adopt their value of $B_{min}=10^8\;{\rm G}$. The magnetic
field decay ceases once the field reaches this value.

\subsubsection{Roche lobe overflow}
\label{roche}

Before the second supernova explosion occurs, the companion of an already formed
pulsar continues its stellar nuclear evolution. If it fills its Roche lobe
during this period, then mass transfer onto the pulsar can occur and the system
becomes an X-ray binary. Some of the mass  from the system is lost and the rest
is accreted onto the neutron star. In addition to increasing the mass of the
neutron star this has three effects:\\ (I) we assume that the magnetic field
decays.  The physical process explaining this quenching might be the
aforementioned ohmic dissipation (changes of the crust's resistance due to
heating during accretion). Other groups consider the magnetic field burial as
another possible cause of the accretion-induced field decay
\citep{1974SvA....18..217B,
1986ApJ...305..235T,2001PASA...18..421M,2001ApJ...557..958C,
2002MNRAS.332..933C,2004MNRAS.348..661K,2005ApJ...625..957L}.  Another option is
tying the evolution of the magnetic field to changes in the pulsar spin period
through the vortex-fluxoid interactions. The proton vortices which carry the
magnetic field are dragged by the neutron vortices inwards or outwards in the
radial direction when the pulsar is spun-up or spun-down, respectively
\citep{1985Ap&SS.115...43M,1991ApJ...366..261R,1991ApJ...382..576R,
1991ApJ...382..587R,2000ApJ...532..514J}. We generally favour the ohmic
dissipation as we adopt an exponential decay. However we do not try to
distinguish between the three possibilities and only note that the
vortex-fluxoid interactions would have consequences for our propeller modelling
as it depends on the spin change (see \ref{propeller}). We model the magnetic
field change as:
\begin{equation}
B=\left(B_0-B_{min}\right)\cdot\exp\left(-\frac{dM}{\Delta
M_d}\right)+B_{min}\;, \label{bdecay}
\end{equation}
where $dM$ is the accreted mass and $\Delta M_d$ is magnetic field decay mass
scale, and $B_{min}$ is the minimal magnetic field of a neutron star.  This
approach is similar to the one adopted by \citet{2008MNRAS.388..393K}.  Other
possibilities based on the observed correlation between the magnetic field and
the estimated accreted mass for binary X-ray sources have also been explored by
\citet{1989Natur.342..656S}.  However, their formula gives very similar results
in the relevant range of accreted mass.  After the mass accretion induced decay,
the field stops to decrease in time, i.e. it is fixed at the final level unless
another mass transfer will happen.  In the latter case the field decays in the
same way as in the first event.  The lower limit for the magnetic field still
applies with $B_{min}=10^8\;{\rm G}.$ \\ (II) the pulsar in the accretion phase
does not emit in the radio. This is because the magnetosphere is filled by the
accreted gas, and the particle acceleration is not effective.  The pulsar is
visible in the X-ray band, but in this work we do not model this.  During the
accretion pulsars will not contribute to our radio detectable population.
\\(III) the spin period of a pulsar is affected. Due to angular momentum
transfer the neutron star tends to corotate with the accreted matter at the
Alfven radius:
\begin{equation}
R_A=\left(\frac{8R^{12}B^4}{M\dot{M}^2G}
\right)^{\frac{1}{7}}\;. 
\end{equation}
The accretion rate $\dot M$ is calculated within  the S\textsc{tar}T\textsc{rack} code   by
considering the detailed evolution of the companion \citep{2008ApJS..174..223B}.
The orbital velocity at Alfven radius equals:
\begin{equation}
\Omega_A=\sqrt{\frac{GM}{R_{A}^3}}\;. 
\end{equation}

Following the accretion, the pulsar will spin with the angular velocity
$\Omega_f=\Omega_A$, where $\Omega_f$ is the final velocity after the mass
transfer.  Depending on the model, it might exhibit different behaviour, as
described in \ref{partial} and \ref{propeller}.

\subsubsection{Partial spin-up}
\label{partial}

If the amount of accreted matter during the Roche lobe overflow (see
\ref{roche}) is sufficient, namely $dM \gid 0.1 M_{\sun}$, we then assume that
the pulsar will be spun up fully to the orbital frequency of the material at the
Alfven radius $\Omega_A$. Note that if the matter at the Alfven radius has an
angular velocity smaller than the neutron star has, it can spin down the neutron
star. This is happening via the propeller effect (see \ref{propeller}).  If the
accreted mass is smaller than $0.1M_{\sun}$ then the pulsar can be only spun-up
to a fraction of $\Omega_A$. We assume that the amount of angular momentum
transferred is proportional to the accreted mass:
\begin{equation}
\Delta J \propto \frac{dM}{0.1 M_{\sun}}\;,
\end{equation}
This assumption yields a linear relationship between the amount of the accreted
mass and the final angular velocity of the pulsar:
\begin{equation}
\Omega_f=\Omega_0+(\Omega_A-\Omega_0)\frac{dM}{0.1 M_{\sun}}\;,
\end{equation}
where $\Omega_f$ is the final angular velocity, $\Omega_A$ is the angular
velocity at the Alfven radius and $\Omega_0$ is the initial angular velocity
before recycling.

\subsubsection{Common envelope}
\label{common_envelope}

The formation of double neutron stars involves quite frequently an unstable mass
transfer, i.e. a common envelope phase. During the common envelope phase the
binary orbit is tightened, and a significant amount of matter from the companion
is expelled. At the same time some matter is accreted onto the neutron star, see
Section \ref{astand} and references therein.  In the S\textsc{tar}T\textsc{rack} code the common
envelope is treated following the \citet{1984ApJ...277..355W} formalism and
assuming that the amount of matter accreted by the neutron star lies in the
range $0.05M_{\sun} \lid dM \lid 0.10M_{\sun}$, randomly drawn from it with a
flat distribution. The duration of the common envelope phase is highly
uncertain, but accretion of such an  amount of matter would correspond to a
highly super-Eddington accretion rate.  At this stage the  angular momentum
transfer is chaotic \citep{1997ApJ...478..723B}, and  we assume that the spin of
the neutron star is not affected on the average by  accretion during this phase,
i.e. the spin period of the pulsar is the same as before the common envelope.
However, the accreted matter quenches the magnetic field of the neutron star,
according to Equation~(\ref{bdecay}).  Thus, a system after a common envelope
phase has a tightened orbit, and the pulsar is more massive with a decreased
magnetic field yet its spin frequency is unchanged.  Another possibility is that
there is no accretion of matter onto the neutron star in that phase.  In these
models the pulsars mainly evolve along the field decay lines, which does not
yield a good fit to the observed population (see Section \ref{popnappdot}).

\subsubsection{The propeller effect}
\label{propeller}

Here we consider the case of  a pulsar   rotating with a greater velocity than
the matter orbiting at the Alfven radius. \citet{Illarionov:1975ei} suggested
that in this case the  propeller effect takes place: the centrifugal force at
the Alfven radius is pushing the matter away, inhibiting the accretion onto the
neutron star, assuming that most of the mass in the disc is ejected from the
system. X-ray observations support the propeller effect (e.g.
\citet{1997ApJ...482L.163C}).

In Section \ref{roche} we describe how accretion can spin up or down the neutron
star.  This propeller mechanism is responsible for spinning down the neutron
star.  However, there has been recent suggestions that the propeller effect
might be ineffective at low accretion rates
\citep{2004ApJ...616L.151R,2010MNRAS.406.1208D}.  The mass cannot be completely
ejected from the system and piles up at the edge of the disc.  After some time
the pressure will overcome the centrifugal barrier and the accretion will
proceed. This happens quasi periodically. We consider the possibility that the
mass can be accreted in the end and that the magnetic field is quenched during
the quasi-periodic accretion events which have no net effect on the pulsar's
spin period.  Our motivation for including this approach is that we need to be
able to reduce the pulsar's magnetic field without affecting its spin period in
order to get a good fit to observations.  Models with this ineffective propeller
treatment are noted with the letter P.

In the remaining models we allow the propeller effect to take place and the
neutron star to be both spun down or  up. The outcome depends  on the value of
the magnetic field and the accretion rate. 

\subsubsection{Evolution of the orbit}

When both pulsars are formed and there is no more mass exchange between the two
neutron stars, the only change in the orbit can be attributed to the emission of
gravitational waves.  This effect is very weak in general, but   the tight
binaries will be strongly influenced by the emission of the gravitational waves.
We calculate the corresponding evolution of the orbit in the first
post-Newtonian approximation.  The time derivatives of the eccentricity and the
semi-major axis are \citep{PhysRev.136.B1224}:
\begin{equation}
\frac{da}{dt}=-\frac{64}{5}\frac{G^3(M_1+M_2)^2}{c^5a^3(1-e^2)^{\frac{7}{2}}}\left[1+e^2\left(\frac{73}{24}+\frac{37}{96}e^2\right)\right],
\end{equation}
\begin{equation}
\frac{de}{dt}=-\frac{304}{5}e\frac{G^3(M_1+M_2)^2}{c^5a^4(1-e^2)^{\frac{5}{2}}}\left(1+\frac{121}{304}e^2\right).
\end{equation}
The evolution of the orbit is taken into account for the entire evolution of
the binary, which is until they coalesce, unless it takes more than  $10\;{\rm
Gyr}$.

The pulsars can be radio loud (see \ref{radiolum}) until the magnetospheres
join. This is when the semi-major axis of the system is smaller than the sum of
light cylinder radii:
\begin{equation}
a<\frac{c}{2}\left(\frac{1}{\Omega_1}+\frac{1}{\Omega_2}\right),
\end{equation}
where $a$ is the semi-major axis, $c$ is the speed of light, $\Omega_1$ and
$\Omega_2$ are angular velocities of the first and second pulsar respectively.
The magnetospheres touch in practice always at the same numerical step as the
coalescence of the pulsars.  This means that they can be radio loud until the
very end.

\subsubsection{Radio luminosity}
\label{radiolum}

In order to check the influence of the selection effects and compare the
populations visible in the gravitational waves and in the radio, we need to know
the radio luminosity of all the pulsars. We use the \citet{1990ApJ...352..222N}
model fitted to the observations, also used by \citet{1997A&A...322..477H}:
\begin{equation}
\label{lum_eq} \log_{10}
L_{400}=\frac{1}{3}\log_{10}\left(\frac{\dot{P}_{-15}}{P^3}\right)+1.635,
\end{equation}
where $P_{-15}=\frac{\dot{P}}{10^{-15}}$.  This luminosity is given in units of
$mJy\times kpc^2$ for observations at $400\; {\rm MHz}$.

Even now, with the bigger sample of pulsars this model provides a reasonably
good description of the observed luminosities when compared to the Australia
Telescope National Facility pulsar
catalogue\footnote{http://www.atnf.csiro.au/research/pulsar/psrcat}
\citep{2005yCat.7245....0M}. Although the spread in the observed luminosities is
clearly visible, the luminosities are clustered around the modelled values.
However selection effects can affect the observed luminosity model, so the
observed luminosities should be compared to a model after taking the selection
effects into account.

An alternative luminosity model has been derived by \citet{2006ApJ...643..332F}.
This model is able to explain the lack of the pile-up of pulsars near the death
lines without invoking the decay of magnetic field. The comparison sample used
by \citet{2006ApJ...643..332F} is very different from our sample. The known DNS
systems contain mainly mildly recycled pulsars.  At the same time the luminosity
model from \citet{2006ApJ...643..332F} is biased towards millisecond pulsars.
The observed DNS population does not contain such objects, so any luminosity
model favouring these pulsars will be found less likely in our likelihood
analysis (see Section \ref{popnappdot}). The choice of luminosity function is
connected to the field decay timescale (see the discussion in Section
\ref{roche}).  For comparison, we present one model with the luminosity function
used  by \citet{2006ApJ...643..332F}:
\begin{equation}
\label{lum_eqfg}
\log_{10}L_{1400}=\log_{10}\left(L_{0}P^{-1.5}P_{15}^{0.5}\right)+L_{corr},
\end{equation}
where $L_{1400}$ is the luminosity at $1400\;{\rm MHz}$, $L_0=0.18\;{\rm mJy
\times kpc^2}$ is the scaling factor and $L_{corr}$ is randomly chosen from
zero-centred normal distribution with standard deviation
$\sigma_{L_{corr}}=0.8$.

The pulsars cease to emit in the radio once they cross the so-called death
lines. After crossing these lines on the $P-\dot{P}$ diagram the pulsar's
emission mechanism fails as electron-positron pairs can no longer be created in
the magnetic field. In this work we assume that all the pulsars cease their
emission after crossing the two death lines defined by
\citep{1994MNRAS.267..513R}:
\begin{equation}
\log_{10}\dot{P}=3.29\times\log_{10}P-16.55\;,
\end{equation}
\begin{equation}
\log_{10}\dot{P}=0.92\times\log_{10}P-18.65\;.
\end{equation}
These relations are empirical. Although there are cases known in which a pulsar
is found beyond these death lines \citep{1999Natur.400..848Y}, they describe the
cut-offs on the $P$-$\dot P$ diagram quite well. None of the DNS systems is
found to be beyond the death lines we adopt here.

For the discussion of the implemented radio selection effects we refer the
reader to Section \ref{radiodet}.

\subsubsection{Numerical step size}

In order to choose the length of the numerical step size during the evolution of
a DNS we calculate the number of time derivatives describing the rate of change
of pulsar properties. The step is chosen to satisfy the following conditions:
(i) the pulsar period changes by less than 1\% (ii) the semi-major axis of the
orbit changes by less than 1\%, (iii) the step is limited by  the beginning of
the next mass transfer episode, (iv) the time step is shorter than $1\;{\rm
Gyr}$.  The last condition was introduced to follow the evolution of mildly
recycled millisecond pulsars with a weak field on wide orbits.

\begin{table}
\caption{Free parameters in the evolution model}
\label{wolpar}
 \begin{tabular}{|c|c|}
 \hline
Parameter& Default value\\
\hline
$\tau_d$	& $5\;{\rm Myr}$\\
$\Delta M_d$	& $0.025\;M_{\sun}$ \\
$B_0$	& $[10^{11}\;{\rm G},10^{13}\;{\rm G}]$\\
$B_{min}$& $10^8\;{\rm G}$ \\
$I$	& $10^{45}\;{\rm g\times cm^2}$ \\
$R$	& $10\;{\rm km}$ \\
$\alpha$	& $30\degr$\\
$\Delta M_{min}$	& $0.1\;{\rm M_{\sun}}$ \\
\hline
\end{tabular}

\medskip $\tau_d$ is the timescale of field decay, $\Delta M_d$ is the mass
scale of field decay, $B_0$ is the initial magnetic field, $B_{min}$ is the
minimal magnetic field, {\em I} is the moment of inertia, {\em R} is the pulsar
radius, $\alpha$ is the angle between rotation and magnetic axis, while $\Delta
M_{min}$ is the accreted mass limit for full spin-up.

\end{table}

\subsubsection{Free parameters}

We list the parameters describing the evolution of the pulsars in
Table~\ref{wolpar}.  The values of the evolutionary parameters for the several
considered models are given in Table \ref{modele}.  The models are denoted as
follows: the first letter in the name of model refers to the stellar evolution
model in the S\textsc{tar}T\textsc{rack} code; the letter P means that the model incorporates
the inefficient propeller effect. The letter F implies that full spin-up is
always possible in the given model. In other words, all the pulsars are recycled
to the orbital velocity at Alfven radius regardless of the amount of accreted
matter; the letter D followed by a number refers to the magnetic field decay
mass scale $\Delta M_d$.  The letter C means that there is no accretion in the
CE phase. The letter I stands for a different assumption about the initial spin
periods, see Section \ref{init}. The letter L is used for models with a
different radio luminosity prescription, see Section \ref{radiolum}. Finally,
the letter T followed by a number gives the value of the magnetic field decay
timescale in Myrs. In the case of T1k and T2k (time decay timescale of 1 and 2
Gyrs) it also means that the magnetic field decays for the lifetime of the
pulsar - the quenching does not stop after an accretion event.

\subsubsection{Motion in the Galaxy}
\label{motion}

In order to calculate the observed radio fluxes of our pulsar population we need
to model their motion in the Milky Way. Therefore we consider a simple model of
the Galaxy consisting of the following three components: a bulge, a disk, and a
halo. The bulge and disk potential are described by the
\citet{1975PASJ...27..533M} type potential, also used by
\citet{1990ApJ...348..485P,1999MNRAS.309..629B}:
\begin{equation}
\Phi(r,z)=
\frac{GM}{\sqrt{R^2+(a+\sqrt{z^2+b^2})^2} } \, ,
\end{equation}
where $M$ is the total mass of a given component, $R=\sqrt{x^2+y^2}$, and $a,b$
are the free parameters.  The halo is described by the density distribution
$\rho=\rho_c [1+(r/r_c)^2]^{-1}$ with a cut-off at $r_{cut}=100$\,kpc above
which the halo density is zero. The corresponding potential for $r<r_{cut}$ is
\begin{equation}
\Phi(r)= - \frac{GM_h}{r_c} \left[ \frac12 \ln\left(1+ \frac{r^2}{r_c^2} \right)
+ \frac{r_c}{r} \arctan\left(\frac{r}{r_c} \right) \right]. 
\end{equation}
We use the following values of the parameters \citep{1991ApJ...381..210B}
describing the bulge (index 1) and disk (index 2)  potential: $a_1=0$\,kpc,
$b_1=0.277$\,kpc, $a_2=4.2$\,kpc, $b_2=0.198$\,kpc, $M_1=1.12\times 10^{10}
\,M_{\sun}$, $M_2=8.78\times 10^{10} \,M_{\sun}$, while for the halo potential
we use $M_h=5.0\times 10^{10} \,M_{\sun}$, and $r_c=6.0$\,kpc.  The distribution
of stars in the disk is assumed to be that of a young disc
\citep{1990ApJ...348..485P}. The radial and vertical distributions are
independent i.e.  the distributions factor out:
\begin{equation} P(R,z)\propto
R(R)dR \, p(z) dz \, ,
\end{equation}
where the radial distribution is $p(R) \propto \exp(-R/R_{exp})$, and $R_{exp}=
4.5\,$kpc, and we introduce an upper cut-off at $R_{max}=20$\,kpc. The vertical
distribution is exponential $p(z) \propto \exp\left(-z/75{\rm pc}\right)$.
  
  We distribute our modelled binaries in the Galaxy by choosing their location
randomly to mimic the densities described above. The velocity of the binary
corresponds to its location in the Galaxy.  We also give both neutron stars
natal kicks, as obtained from the S\textsc{tar}T\textsc{rack}. We use the leapfrog method in
the KDK (kick-drift-kick) scheme with a constant ($10^5$ years) time step.  This
approach is fully symplectic. In other words, it is equivalent to solving the
Hamilton's equations of motion and fully preserves energy and angular momentum
(e.g. \citet{2010AJ....139..803Q}).

\begin{table*}
\begin{minipage}{120mm}

 \caption{Models of pulsar evolution. The last column describes what is changed
in the model with respect to the standard possibilities, including initial
period variation in range 10 to 100 ms, no accretion in the CE phase, a
different luminosity law and a continuous decay of magnetic field.}

\label{modele}

\begin{tabular}{|c|c|c|c|c|c|c|}
\hline
Model & $\tau_d$ & $\Delta M_d$ & propeller & spin-up & S\textsc{tar}T\textsc{rack} model & other\\
\hline
A    & 5 & 0.025 & yes & partial possible & A & - \\
AF   & 5 & 0.025 & yes & always full  & A & -  \\
APD05& 5 & 0.05 & inefficient & partial possible & A & -  \\
APD05L  & 5 & 0.025 & inefficient & always full  & A & luminosity  \\
APD05I & 5 & 0.05 & inefficient & partial possible & A & initial period \\
APD05T1k & 1000 & 0.05 & inefficient & partial possible  & A & magnetic field \\
APD003T2k & 2000 & 0.0033 & inefficient & partial possible & A & magnetic field \\
AP   & 5 & 0.025 & inefficient & partial possible & A & -  \\
APC & 5 & 0.025 & inefficient & partial possible & A & no CE \\
APT20& 20& 0.025 & inefficient & partial possible & A & - \\
HP   & 5 & 0.025 & inefficient & partial possible & H & -  \\
SP   & 5 & 0.025 & inefficient & partial possible & S & -  \\

\hline
\end{tabular}
\end{minipage}
\end{table*}

\subsection{Detection in the radio band}
\label{radiodet}

We use the radiometer equation \citep{1985ApJ...294L..25D} to calculate the
minimum flux necessary to detect the pulsar for a given signal-to-noise (S/N)
threshold:
\begin{equation}
\label{Smin_eq} S_{min} = \beta
\frac{(S/N_{min})\left(T_{rec} + T_{sky}\right)}{G\sqrt{n_{p}t_{int}\Delta
f}}\sqrt{\frac{W_{e}}{P-W_{e}}}\;,
\end{equation} 
where $\beta$ is a parameter arising from digitisation errors (e.g. $\beta=1.25$
for one bit digitisers) and other effects reducing the $S/N$, such as radio
frequency interference or distortion of the bandpass, $T_{rec}$ is the receiver
noise temperature, $T_{sky}$ is the sky temperature in the direction of given
pulsar, G is the antenna gain, $n_p$ is the number of polarisations, $t_{int}$
is integration time, $\Delta f$ is the receiver bandwidth, W is the pulse width
and P is the pulsar period \citep{2004hpa..book.....L}. The first part of
Equation \ref{Smin_eq} is the usual radiometer equation, while the second part
(the square root) is the pulsar term.   We will assume that $\beta=1$.
Other values required to calculate the minimum flux for the pulsar's detection
are taken to mimic the most successful pulsar survey, the Parkes Multibeam
Pulsar Survey (e.g. \citet{2001MNRAS.328...17M}). 

We have taken into account the broadening of the pulse width due to the
distribution of electrons in the Galaxy \citep{2002astro.ph..7156C},
interstellar scattering and sampling time of a survey, so that the effective
width of the pulse entering Equation (\ref{Smin_eq}) is (e.g.
\citet{2006MNRAS.368..283B}):
\begin{equation} W_{e}^2=W_{i}^2+\tau_{samp}^2 +
\left(\tau_{samp}\cdot\frac{DM}{DM_0}\right)^2 +\tau_{scatt}^2\;,
\end{equation}
where $W_{i}$ is the intrinsic pulse width, $\tau_{samp}$ is the survey sampling
time, DM is the dispersion measure in the direction of the pulsar, $DM_0$ is the
diagonal dispersion measure of the survey and $\tau_{scatt}$ is the interstellar
scattering time. \citet{2004ApJ...605..759B} have obtained a fit of
$\tau_{scatt}$ as a function of the dispersion measure and this model is used in
this paper. We assume that all the pulsars have the duty cycle
$\frac{W_i}{P}=0.05$.  Observations show that there is a dependence of the duty
cycle on the pulsar's period (e.g. \citet{1988MNRAS.234..477L}), but we neglect
that for simplicity. Effective widths, dispersion measures and $S_{min}$ are
calculated with a modified version of the
PSREVOLVE\footnote{http://astronomy.swin.edu.au/\textasciitilde
fdonea/psrevolve.html} code written at the Centre for Astrophysics and
Supercomputing,  Swinburne University of Technology.

To calculate the limiting  flux density we choose a realistic signal-to-noise
detection threshold of $\left(S/N_{min}\right)\ge10$.  We then compare $S_{min}$
with our modelled luminosity of all the radio loud pulsars.  If the observed
flux exceeds this limit we assume that this pulsar is detectable in the radio.
 
 While calculating the properties of the radio population, we model the beam
width of the pulsar as a function of its spin period. We adopt the results of
\citet{1998MNRAS.298..625T} and assume that the beaming fraction is:
\begin{equation}
f_{beaming}=9\left(\log\frac{P}{10}\right)^2+3,
\end{equation}
where $P$ is the rotational period of the pulsar in seconds. This formula has
been obtained by fitting to slow rotating pulsars and is a good fit for pulsars
with $P \ga 100 ms$.
 
Since the radio luminosity model described in \ref{radiolum} provides the radio
luminosity at $400\; {\rm MHz}$, we need to translate these values into
luminosities at $1420\; {\rm MHz}$, as this is the frequency used by the Parkes
Multibeam Pulsar Survey:
\begin{equation}
L_{1420}=L_{400}^{\alpha_{sp}},
\end{equation}
where $L_{1420}$ is the luminosity at  $1420\; {\rm MHz}$, $L_{400}$ is the
luminosity at $400\; {\rm MHz}$ calculated from Equation \ref{lum_eq}, and
$\alpha_{sp}$ is the spectral index. We adopt the $\alpha_{sp}=-1.8$ value from
\citet{2000A&AS..147..195M}

\subsection{Detection in gravitational waves}
\label{gwdet}

The gravitational wave detectors such as LIGO and VIRGO, (for sensitivity
estimates see: \citep{Flanagan:1997sx} and  \citep{2006A&A...459.1001K}) are
currently working and in principle can detect neutron star coalescence with a
chirp mass of $1.2\;{\rm M_{\sun}}$ as far as 18~Mpc.  The signal-to-noise ratio
from the inspiral phase of two compact objects is given by
\begin{equation}
\frac{S}{N}\sim M_{chirp}^\frac{5}{6}\times\frac{1}{D}\;,
\end{equation}
where $D$ is the distance from the observer and $M_{chirp}=\left(M_1
M_2\right)^{0.6}\left(M_1+M_2\right)^{-0.2}$ is the chirp mass, where $M_1$ and
$M_2$ are the masses of the first and second star in the system respectively.
We define the DNS population observable in the gravitational waves as consisting
of consisting of  binaries with merger time shorter than the Hubble time. All
the observable quantities of the DNSs are weighted by the volume in which they
are observable, i.e.:
\begin{equation}
V \sim M_{chirp}^\frac{5}{2}\;.
\end{equation}
Thus the gravitational wave population is the population residing in multiple
galaxies and we assume that the populations in these galaxies resemble the one
in our Galaxy.  We neglect a possible detection of gravitational waves in the
merger and the ringdown phases of the coalescence.

\section{Results}
\label{results}
\subsection{Example of a binary evolution}
\label{ewolucjaex}

\begin{figure}
 \includegraphics[width=\columnwidth]{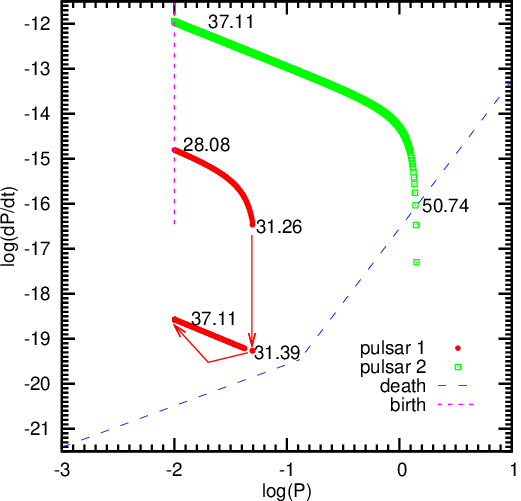}
 \caption[An example of evolution on $P-\dot{P}$ diagram]{An example of a
neutron star binary evolution on the $P-\dot{P}$ diagram.  The dotted line
corresponds to the birth line, where all neutron star are formed in our model.
The solid lines represent the death lines.  The numbers correspond to
characteristic ages in Myrs and are explained in in the text (Section
\ref{ewolucjaex}). Arrows show how the mass transfer affects the first pulsar.}
 \label{pdpex}
\end{figure}

In subsection \ref{ewolucja} we described our phenomenological model of pulsar
evolution. In  Fig. \ref{pdpex} we present an example of an evolutionary path
for one of the simulated binaries. This example is based on the AF model,
founded on the S\textsc{tar}T\textsc{rack} model A with full spin regardless of the total
accreted mass.

The pulsars are born with a rotational period $P_{ini}=10\;{\rm ms}$ at the
birth line (the dotted line on Fig. \ref{pdpex}).  The initial value of the
magnetic field, which determines the spin period derivative, is drawn from a
flat distribution as described in Section \ref{ewolucja}. 

The binary starts on the zero age main sequence at $t=0$. After 28.08~Myrs the
first pulsar is born.  At that time, the system consists of a pulsar and a
massive rejuvenated companion. The neutron star initially evolves in the $P-\dot
P$ plane along the line of constant magnetic field before turning down when the
field decay becomes significant. At $t=31.26\;{\rm Myrs}$ the nuclear evolution
of the companion plunges the system into the common envelope phase.  The
magnetic field of the pulsar is quenched and it falls close to the death line
without changing the spin period. The companion loses its envelope and becomes a
helium star while the neutron star reappears in the radio.  Its $\dot{P}$ is
very small now so that it barely evolves until the second mass transfer occurs.
This mass transfer is a stable Roche lobe overflow and commences at
$t=31.39$~Myrs. The system becomes an X-ray binary and the neutron star is
mildly recycled to a period $\approx 0.01s$.  The amount of accreted matter is
very small ($\sim10^{-5}M_{\sun}$), but in this model full recycling is allowed
regardless of the total accreted mass. In many other models (the ones without F
in their name) we limit the accretion rate to the Eddington limit and
consequently the recycling is weaker. Soon after this mass transfer, at
$t=37.11$~Myrs, the second pulsar is born, and for the following  23~Myrs the
binary contains two radio loud pulsars. The second pulsar will evolve similarly
to the first one. Initially it moves along the constant field line. As a
consequence of our short field decay timescale it turns down and moves almost
vertically. No mass transfers can occur as both stars are past the supernova
explosion. At around $t=50.74\;{\rm Myrs}$ the second born pulsar falls below
the death line. At this stage the system contains a mildly recycled pulsar which
slowly evolves to the death line. It does not reach it throughout the whole
simulation, which ends after 10~Gyrs. Some of the pulsars in our simulations
manage to pass the death line before coalescence.  This binary is relatively
wide, so it does not coalesce within that time.

\begin{figure*}
\includegraphics[width=\columnwidth]{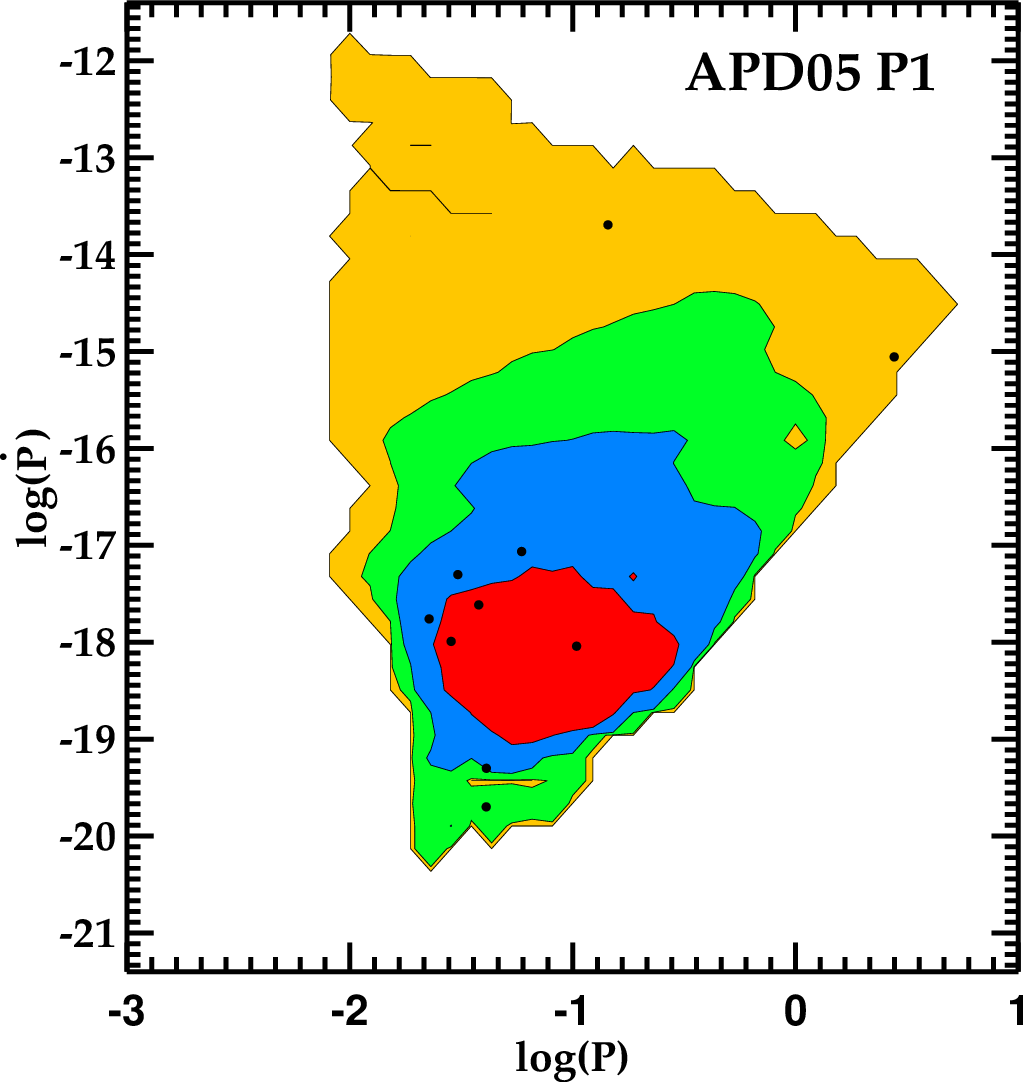}
\includegraphics[width=\columnwidth]{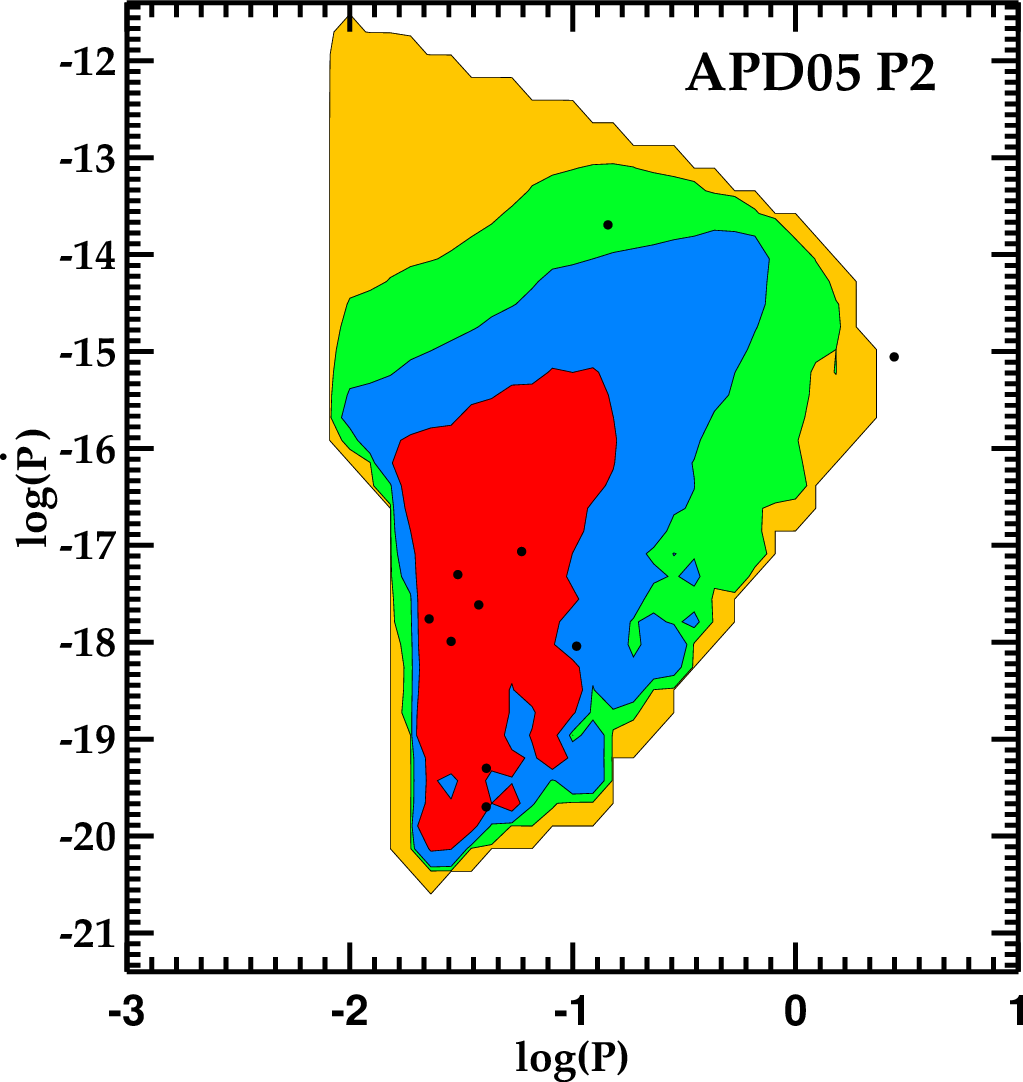}
\includegraphics[width=\columnwidth]{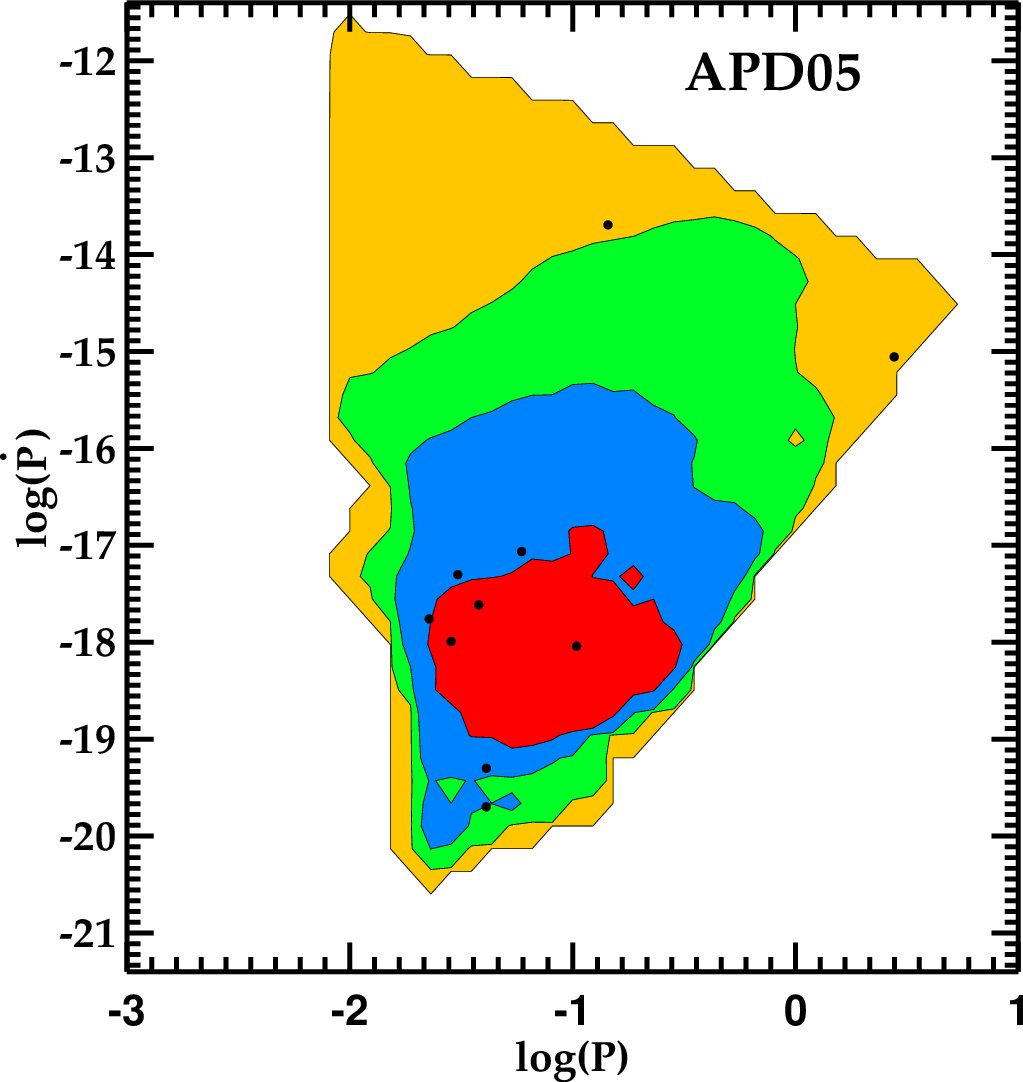}
\includegraphics[width=\columnwidth]{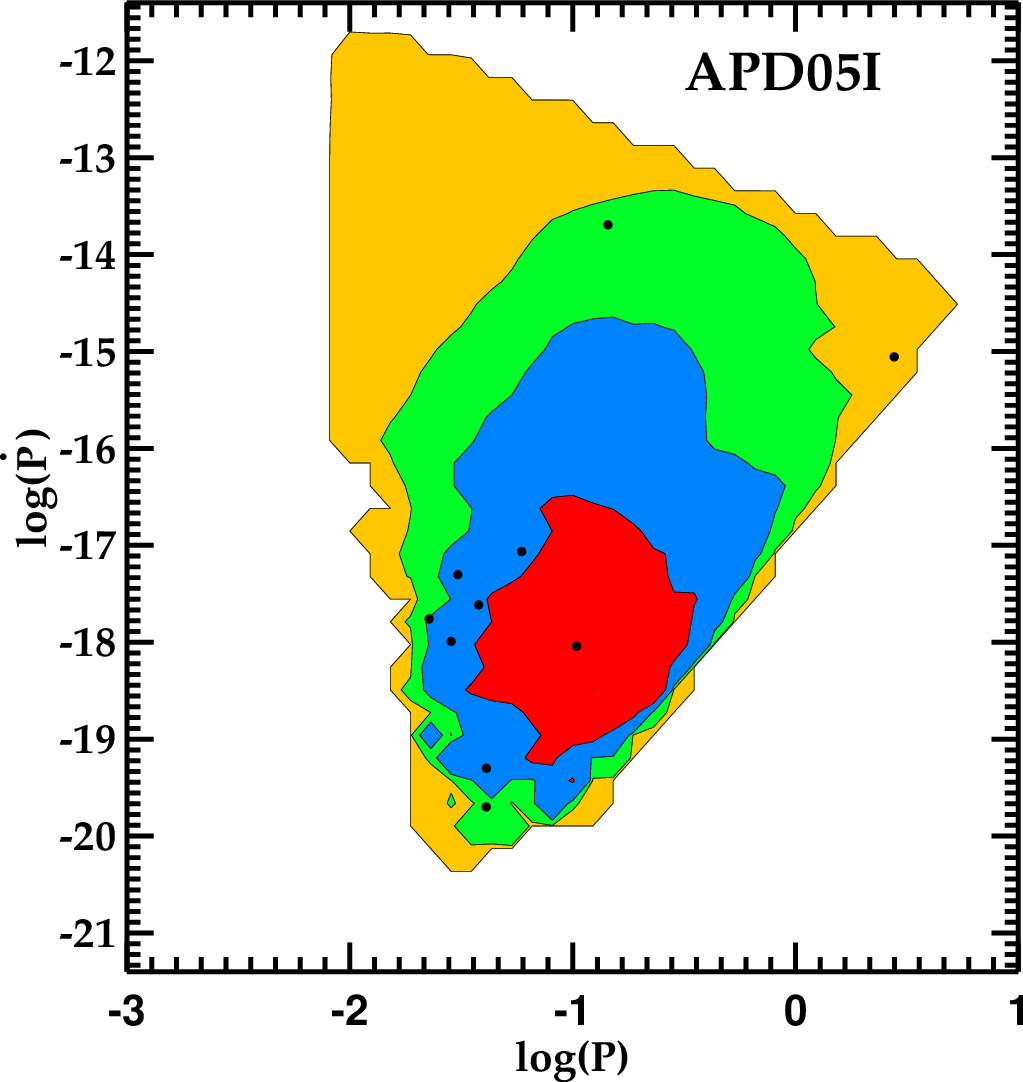}
 \caption{Probability density on $P-\dot{P}$ diagram for APD05 model, for both
first and second born pulsars separately (P1 and P2, respectively), for both
pulsars together and for the APD05I model, where a flat distribution of initial
spin periods is assumed.  The levels correspond to contours containing 68, 95
and 99 percent of the objects. Black points correspond to observations.}
\label{pop2Ar} \end{figure*}

\begin{figure*}
\includegraphics[width=\columnwidth]{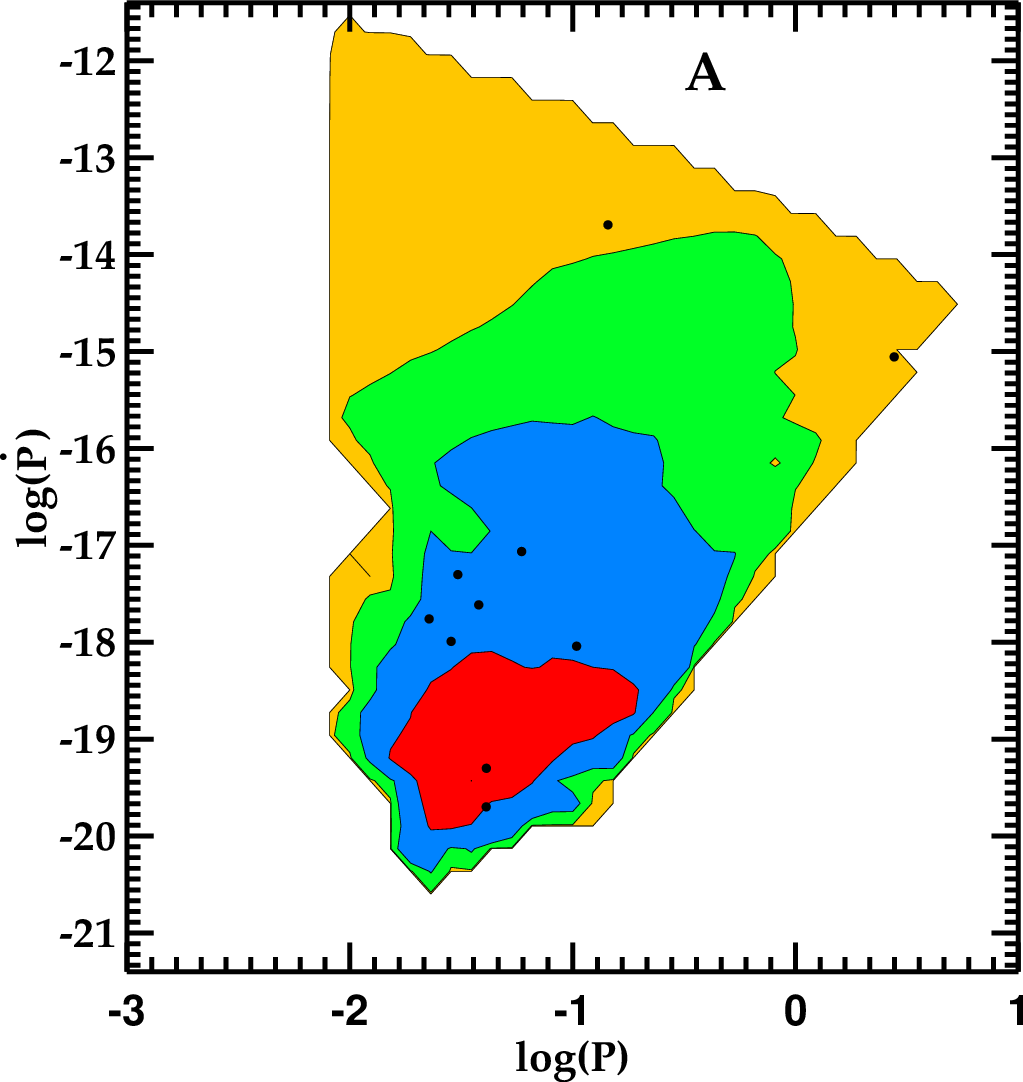}
\includegraphics[width=\columnwidth]{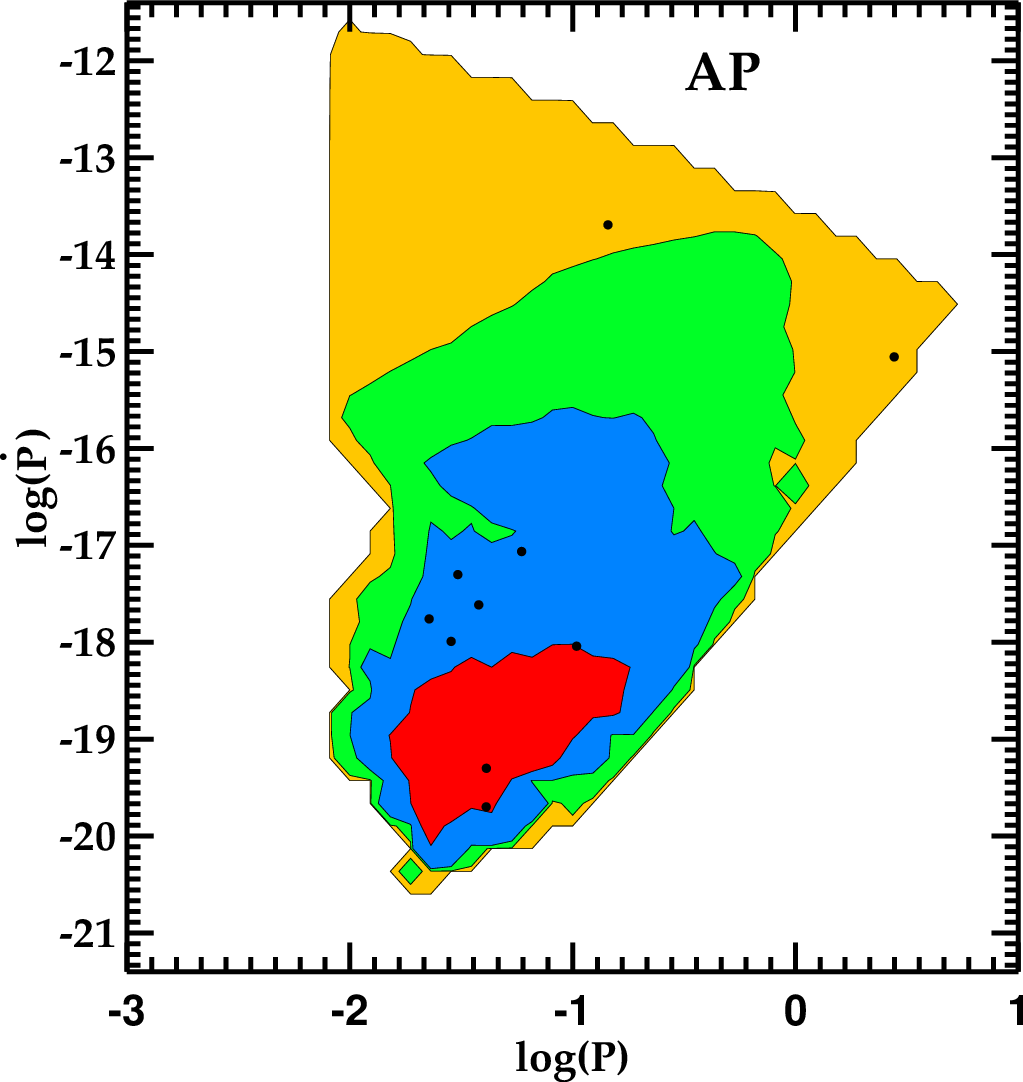}
\includegraphics[width=\columnwidth]{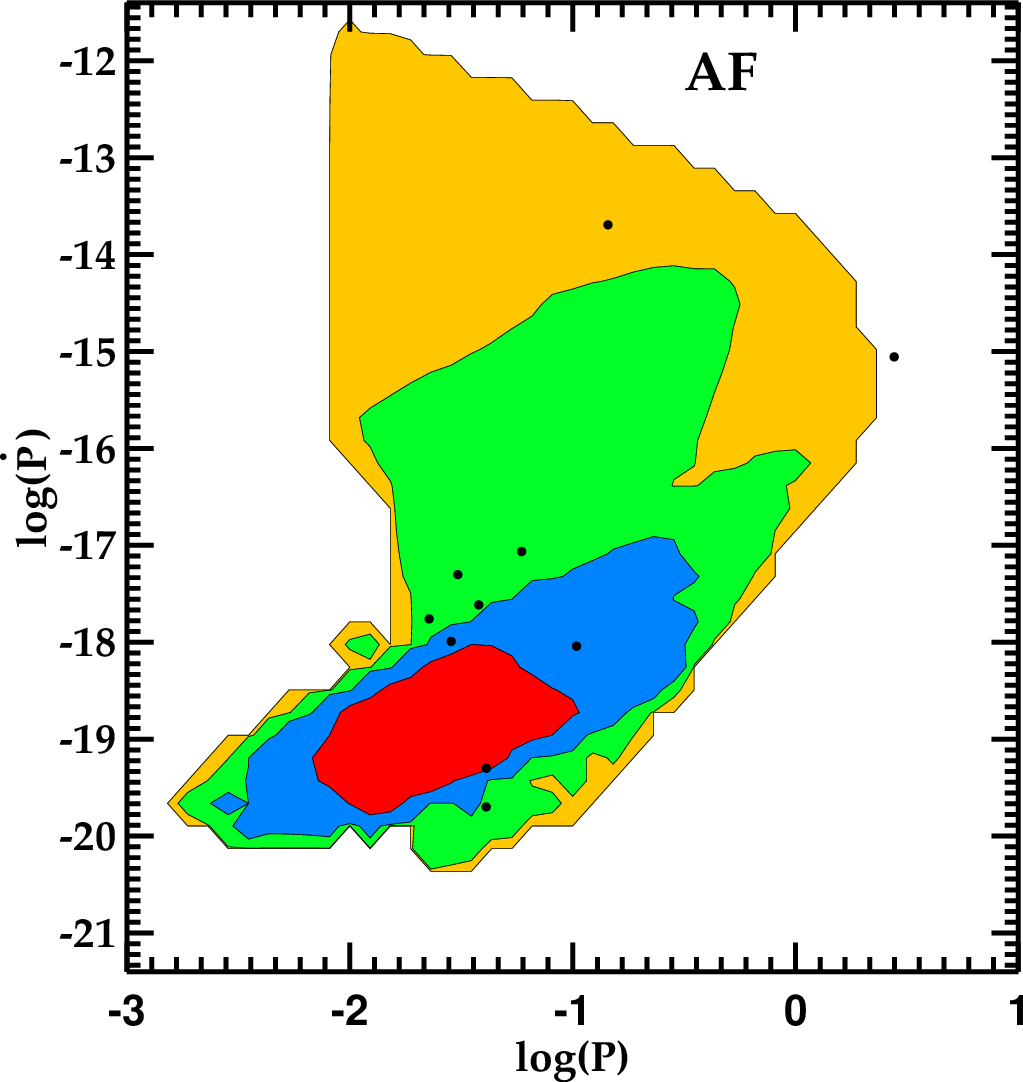}
\includegraphics[width=\columnwidth]{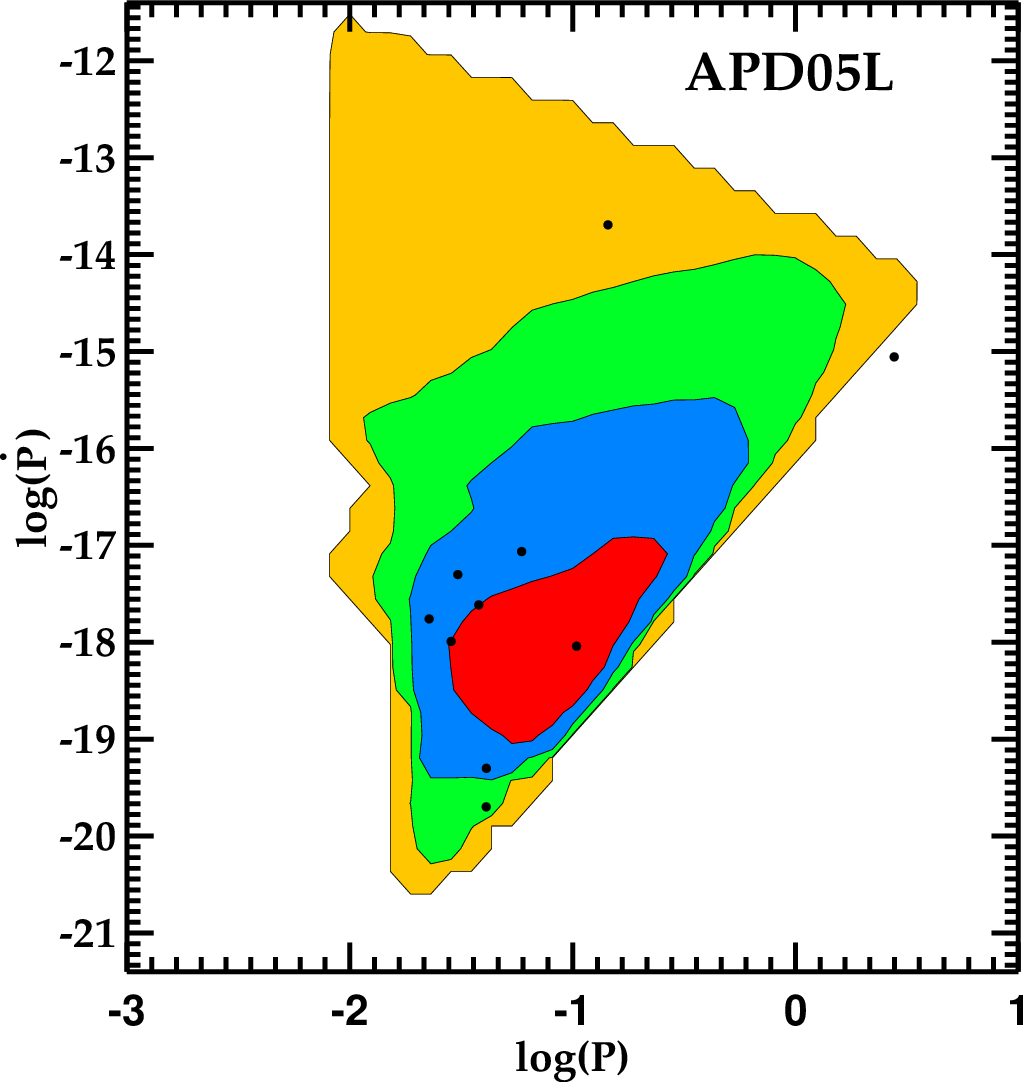}
 \contcaption{Probability density on $P-\dot{P}$ diagram for the A, AP, AF and
APD05L models.  The levels correspond to contours containing 68, 95 and 99
percent of the objects. The outer contour delimits the region containing all the
objects in the simulation. Black points correspond to observations.}
\label{pop2Br} \end{figure*}

\begin{figure*}
\includegraphics[width=\columnwidth]{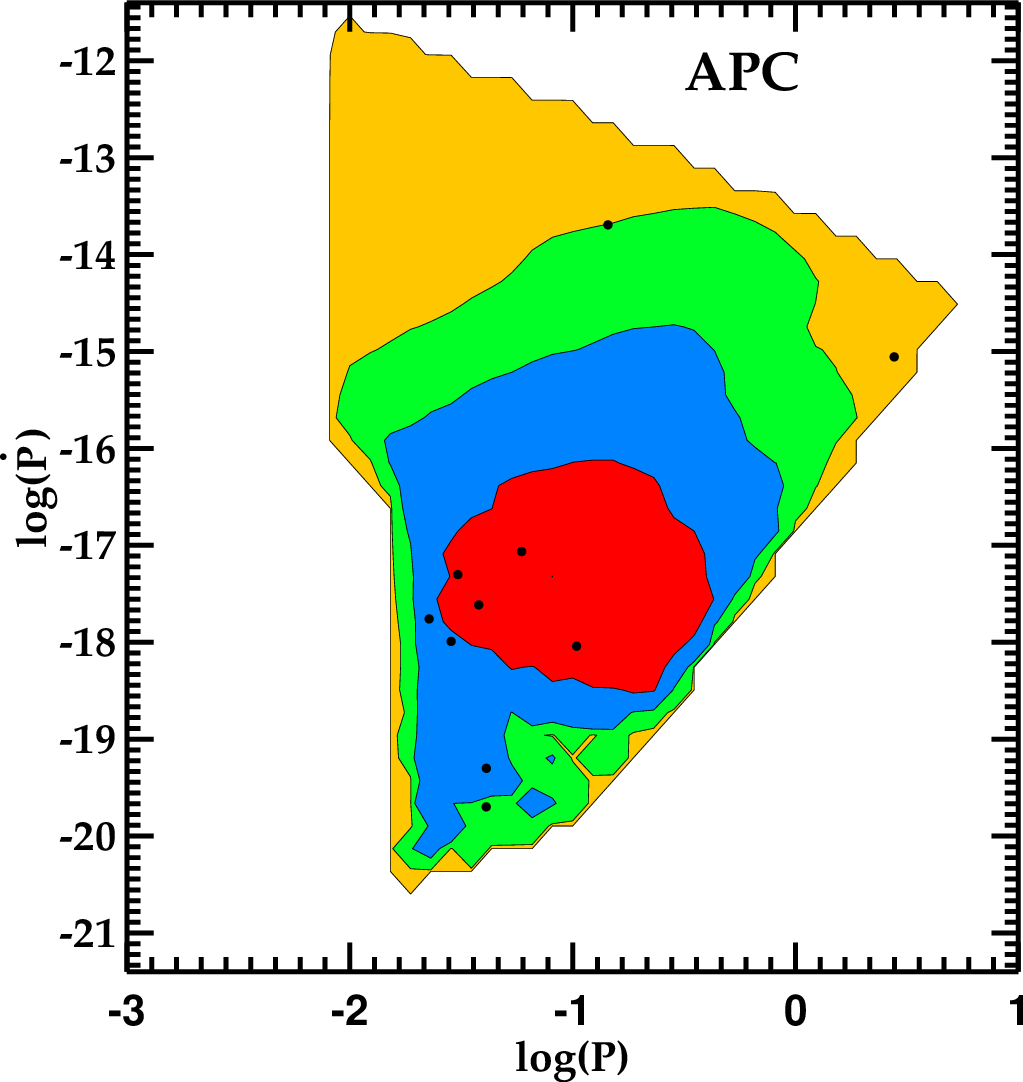}
\includegraphics[width=\columnwidth]{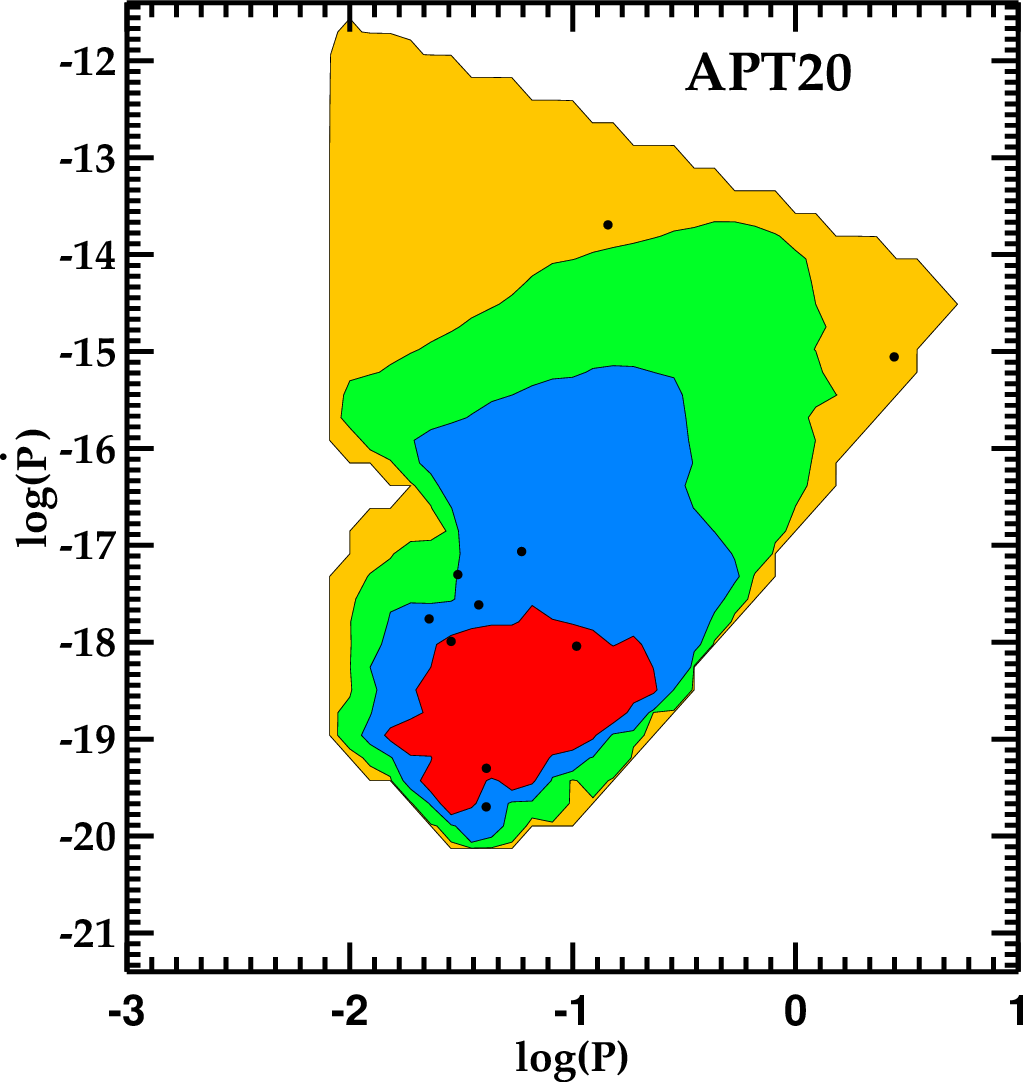}
\includegraphics[width=\columnwidth]{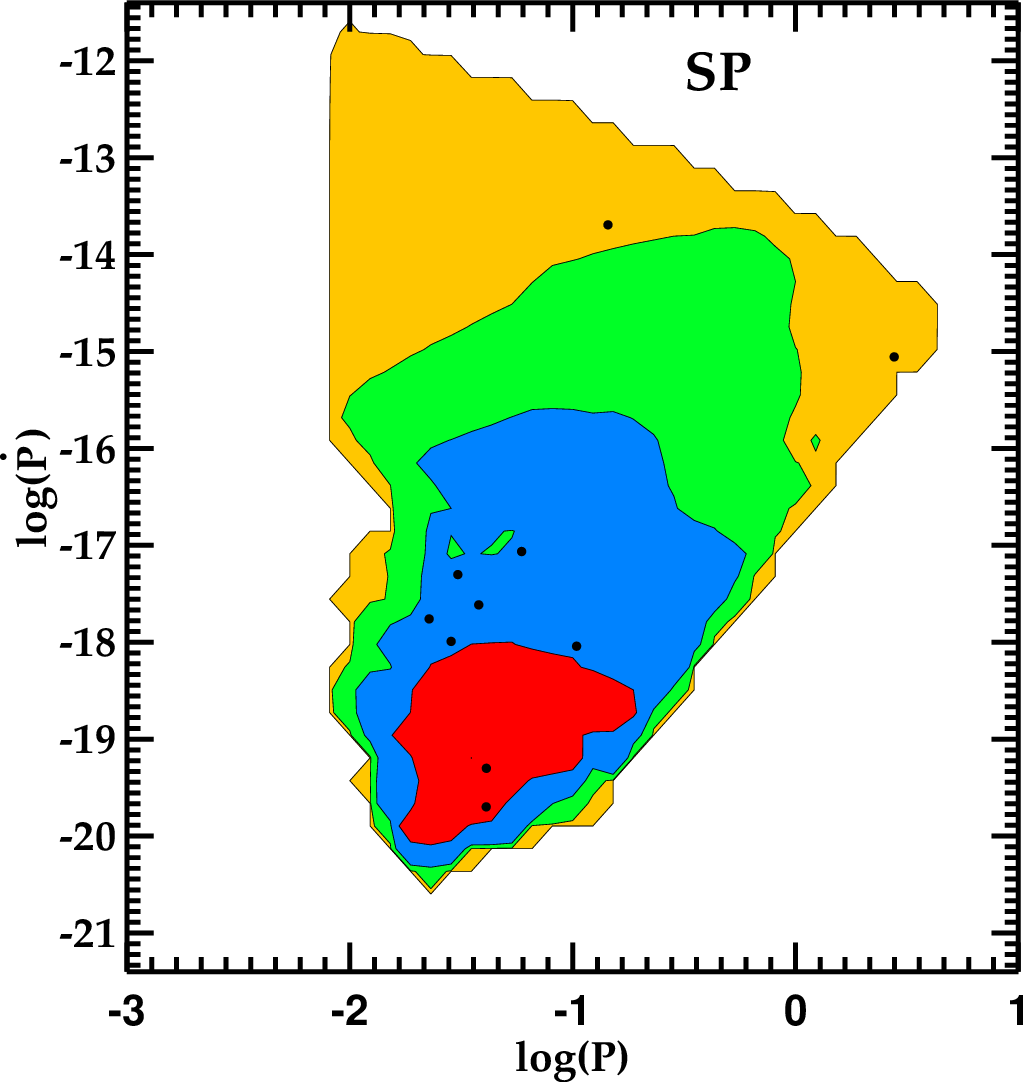}
\includegraphics[width=\columnwidth]{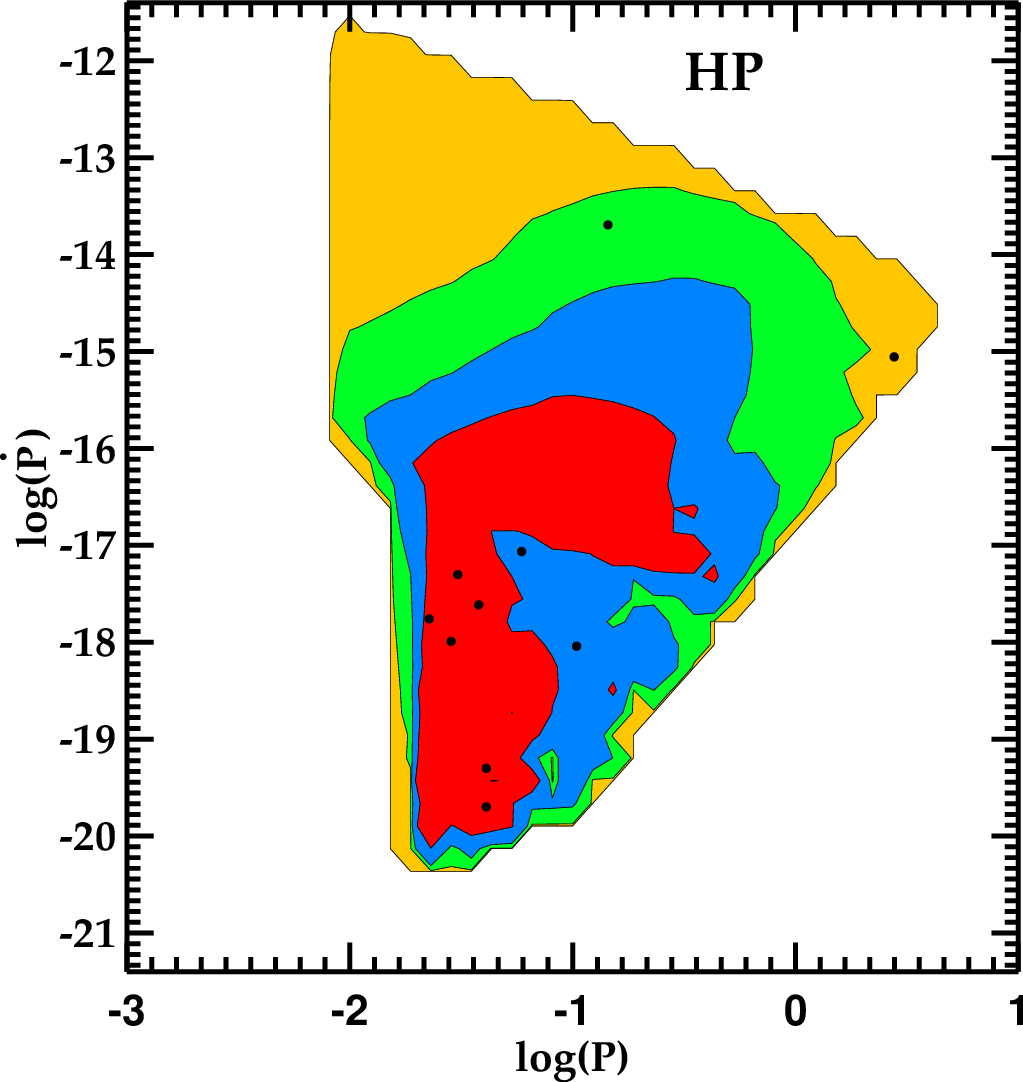}
 \contcaption{Probability density on $P-\dot{P}$ diagram for the APC, APT20, SP
and HP models.  The levels correspond to contours containing 68, 95 and 99
percent of the objects.  Black points correspond to observations.}
\label{pop2Cr} \end{figure*}

\begin{figure*}
\includegraphics[width=\columnwidth]{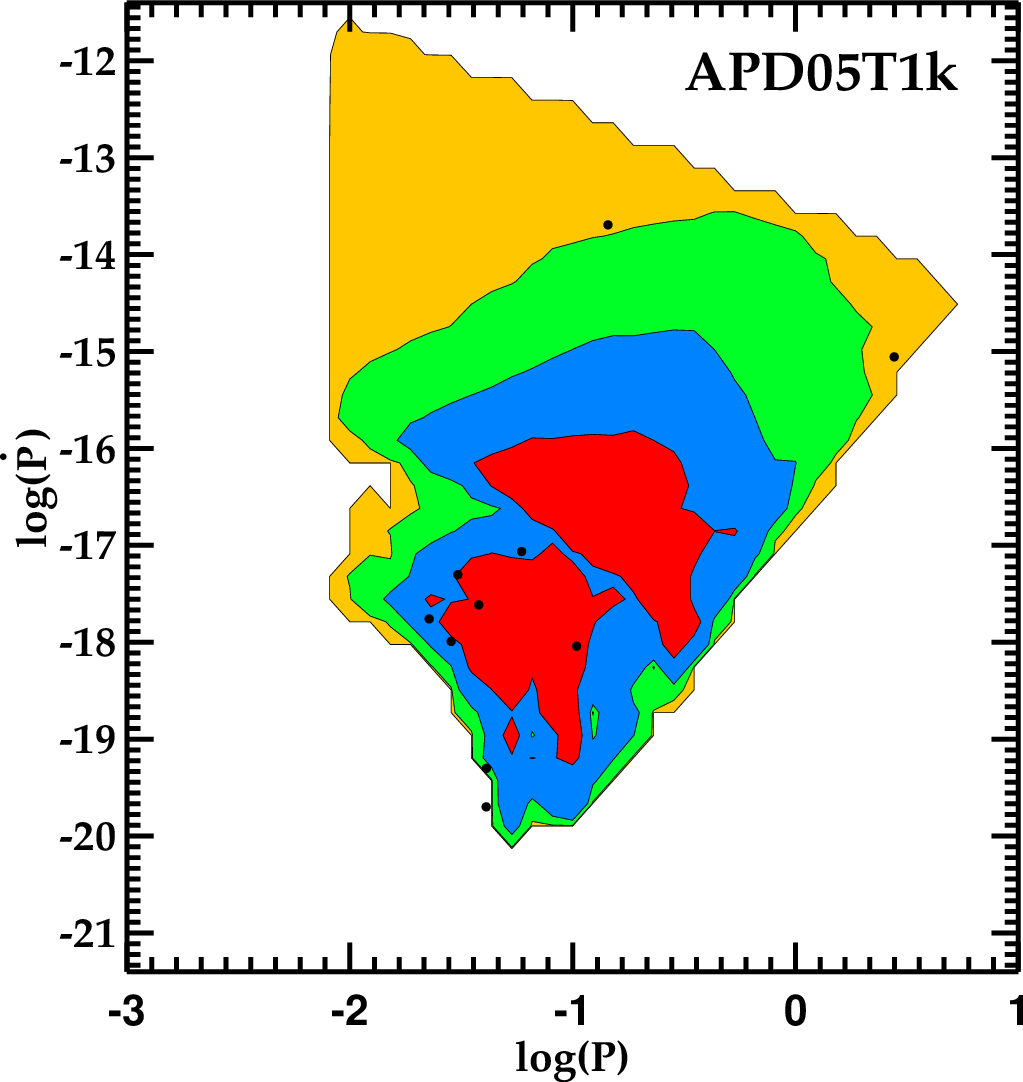}
\includegraphics[width=\columnwidth]{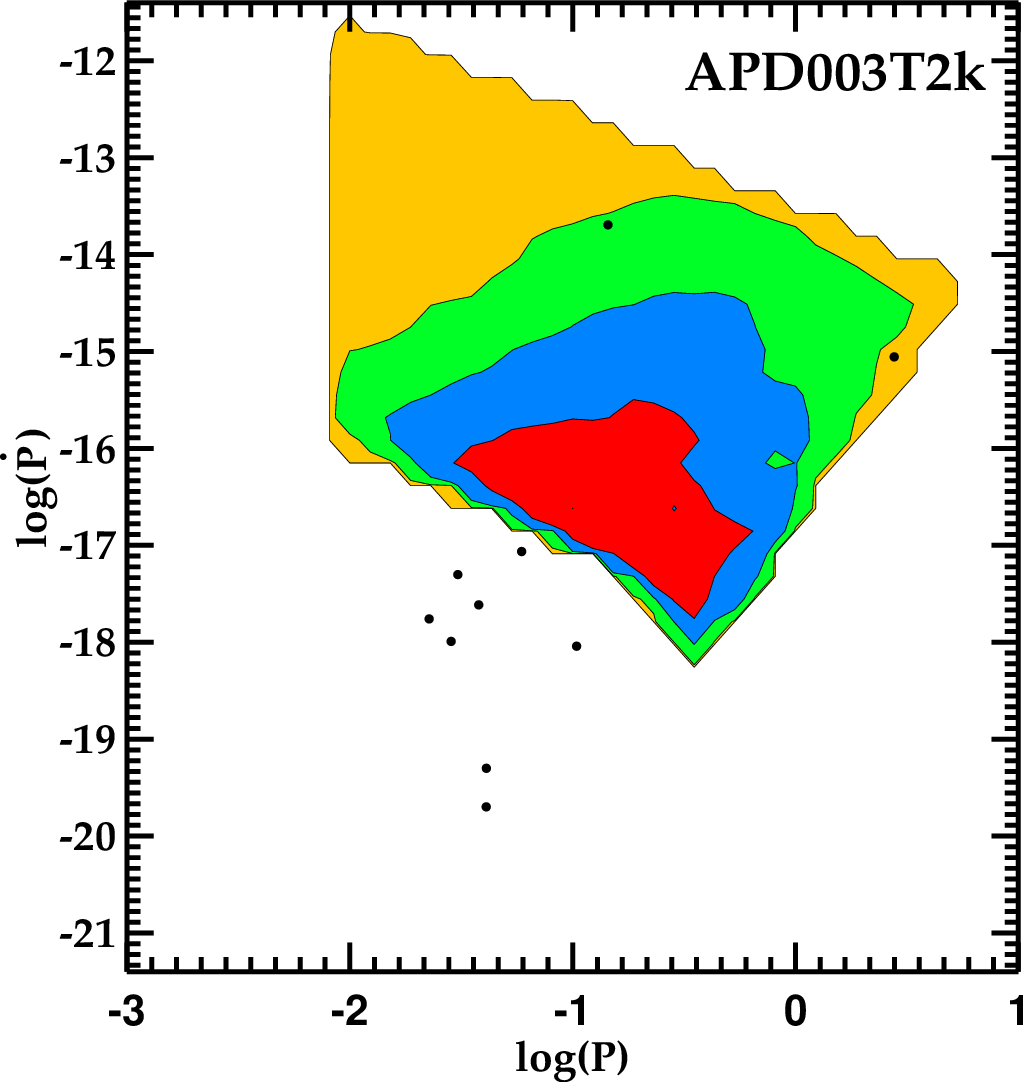}
 \contcaption{Probability density on $P-\dot{P}$ diagram for the APD05T1k and
APD003T2k models.  The levels correspond to contours containing 68, 95 and 99
percent of the objects. The outer contour delimits the region containing all the
objects in the simulation. Black points correspond to observations.}
\label{pop2Dr} \end{figure*}

\subsection{Population on the $P-\dot{P}$ diagram}
\label{popnappdot}

In order to analyse the properties of the population of DNS we have to take the
selection effects into account. We define two types of selection effects that
lead to differences between the intrinsic and observed population. The first
effect is related to the sensitivity of the telescope, the properties of the
pulsar and the ISM along the line of sight. We have modelled these selection
effects to reflect those that affected Parkes Multibeam Pulsar Survey, as
described in Section \ref{radiodet}. Additionally, it has been noted that
binaries with an orbital period between $0.3$ and $4$h are difficult to detect
\citep{Faulkner}. Therefore we remove these systems from the observed
population. The latter effect proves to be insignificant when it comes to the
statistical properties of the observed pulsars, except for their binary periods.

We start with the analysis of the population of DNS systems on the $P-\dot P$
diagram.  To obtain the present observable population of binary neutron stars we
assume that the star formation rate in the galaxy is constant. We position the
binaries in the Galaxy according to the model described in Section~\ref{motion}.
Then we propagate them in the Galactic potential, giving them the additional
kick velocities that they receive from the supernova explosions. We then look
for the observable population taking into account the described selection
effects.  The density of the objects in the $P-\dot P$ plane is the total time
that each pulsar spends in a given cell on the $\log P- \log\dot P$ plane, i.e.
the density in a given cell is :
\begin{equation}
f_{ij}=\frac{F_{ij}}{F}\frac{1}{\Delta\log P\Delta\log\dot{P}}\;, \label{fij}
\end{equation}
where $\Delta\log\left(P/1{\rm s}\right)=0.091$ and
$\Delta\log\left(\dot{P}/1{\rm s^{-1}}\right)=0.261$ denote the bin width of the
spin period and its derivative respectively, $F=\sum\limits_{i,j=1}^{44}F_{ij}$
and
\begin{equation}
F_{ij}=\sum\limits_{p=1}^{N_{ij}}t^p_{ij}f_{beaming}^p,
\end{equation}
where $N_{ij}$ is the number of simulated pulsars with $P$ contained in the i-th
and $\dot{P}$ in j-th bin and $t^p_{ij}$ is the time the p-th pulsar spends in
these bins and $f_{beaming}^p$ is the beaming factor for p-th pulsar at a given
time.  The range of $\log_{10}P$ from $-3$ to $1$ as well as the
$\log_{10}\dot{P}$ range from $-21.5$ to $-11$ were divided into 44 bins.

The density of the population in the $P-\dot{P}$ plane (shown on Fig.
\ref{pop2Ar} and its continuations) is strongly affected by the radio selection
effects.  The properties of first and second born pulsars contribute to this
density. In majority of the models, the first born pulsar will dominate, as it
is mildly recycled and emits longer in the radio. The details of the binary
evolution and the recycling process strongly influence the region where we
expect the recycled pulsars to be.

All the binary systems in each model have been assigned four sets of initial
parameters, the values of which were randomly chosen, in order to check how
much the results can vary. The presented $P-\dot{P}$ diagrams are an average
of these four simulation runs. Removing a small number of systems and adding
another realization of any given model did not affect the results significantly.

The first two plots show the probability densities on the $P-\dot{P}$ diagram
for the first and second pulsar in our best model, APD05 (APD05 P1 and APD05 P2
respectively).  In this model we consider the inefficient propeller effect and
increase the mass scale of accretion induced field decay to $0.05{\rm
M_{\sun}}$.  It is clear that the first born pulsars are concentrated in the
region of mildly recycled pulsars. The pulsars which are born second evolve very
simply. Initially they follow the lines of constant magnetic field and after few
million years they quickly move down. This is because of the short field decay
timescale.  The combined population is shown on the plot denoted by APD05. It
resembles the plot showing the first born pulsar. This is expected, as the
mildly recycled pulsars evolve along the constant magnetic field line. Hence
they spend a long time in the most densely populated region. The total
contribution of the second born pulsar to this model is only $15\%$. The only
problem with this model is that J1518+4904 and J1829+2456 are at the edge of the
low density regions.

The next plot, showing the APD05I model, is again quite similar. The difference
here is that the initial spin periods are chosen randomly in a range from 10 to
100 ms. Introducing the slower spinning pulsars shifts the whole population
towards longer periods. Note that many of the observed pulsars now fall in the
lower density regions. This results in a smaller likelihood of this model (see
Section \ref{porow}). For the other models, we keep the assumption of a constant
$P_{ini}=10\;{\rm ms}$ for each pulsar. Without this assumption it is harder to
recycle our pulsars enough to match the observed population, unless the
accretion physics is altered.

The next four plots present the A, AP, AF and APD05L models. The first one is
the model based on our initial estimates of the evolutionary parameters. It
predicts too many pulsars with very small $\dot{P}$ (of the order of
$10^{-19}$).  Replacing the propeller effect with our modified prescription
improves the situation only slightly, as demonstrated by the AP model. The
highest probability density extends now to somewhat higher, but still too low,
$\dot{P}$ values. The relatively small number of binaries in which the propeller
effect takes place, explains why this change has such a small impact.  Next, we
present a model with full recycling regardless of the amount of matter accreted
- the AF model. In this case, the population of the recycled pulsars extends to
periods below 20 ms.  There are no such pulsars observed in the DNS systems.

As mentioned in the Section \ref{radiolum}, we present the APD05L model. This is
a variation of the APD05 model with the radio luminosity law defined by Equation
\ref{lum_eqfg}. This model looks more like the AP or A model with the contour
containing $68\%$ of the systems moved to slower rotators with higher $\dot{P}$
values, rather than like the APD05 model it originates from. Based on this
model's low likelihood, we decide that the luminosity prescription of
\citet{1990ApJ...352..222N} yields better results in our models.

We then consider what happens when we do not allow any accretion during the CE
phase. The corresponding model, APC, is similar to the APD05 model, but it is
more likely now to find pulsars with higher values of $\dot{P}$.  This directly
follows from lack of accretion in the CE phase. Without the accretion, each
pulsar will have a stronger magnetic field on average. This effect would be even
more pronounced, but the first born pulsars will have more time to quench their
magnetic field, as we allow them to do that until the first accretion takes
place. 

Next we check how the timescale of the spontaneous decay of magnetic field
influences the overall shape of the distribution in the $P-\dot P$ diagram, see
the AP and APT20 model. Changing the timescale to 20 Myrs does not have a big
efffect. The highest density region is concentrated around higher $\dot{P}$ as
the magnetic field is stronger for the majority of pulsars. Another change is
that the pulsars will start moving down vertically when they reach longer spin
periods. This is a direct consequence of the longer decay timescale.

Before considering even longer timescales, we first turn to the two models based
on the different S\textsc{tar}T\textsc{rack} stellar evolution models, namely the SP and HP
models. The SP model only differs from the AP in the way the initial masses are
calculated. As expected, the SP and AP look very much alike. The contribution of
the second born pulsar to the total density in the $P-\dot{P}$ space in this
model is $\sim 12 \%$.  Differences will be visible in the distribution of the
chirp masses and the density in the primary mass versus the mass ratio planes,
see Section \ref{expected_masses}. The HP model includes the hyper-critical
accretion in the CE phase, which leads to many pulsars having the weakest
allowed magnetic field. These pulsars fall below the death line and are no
longer emitting in the radio. During the next accretion event they could in
principle be recycled and become millisecond pulsars, but in the binaries that
give rise to double neutron star this second accretion phase will be too short
and will not spin them up enough. The vertical part of the high density contour
originates mainly from the second born pulsars (compare with APD05 P2). The part
that roughly follows the constant field lines comes from the first born pulsars
that have only experienced a stable mass transfer.  After the accretion phase
they move along the constant field lines.  As much as $68\%$ of  the
contribution to the total density in the $P-\dot{P}$ space comes from the second
born pulsar, as opposed to $\sim 15\%$ for models based on the A S\textsc{tar}T\textsc{rack}
mode or $\sim 12\%$ for the SP model.

Finally, we come back to the question of the timescale of the magnetic field
decay. We present two different models. At first, we show the ADP05T1k model,
which is the same as the APD05, but with $\tau_{d} = 1000{\rm Myr}$ and a field
decay occurring throughout the lifetime of a pulsar. The contour containing 68
\% of the total density is split into two separate regions. One of them, at high
$\dot{P}$ values, is populated by the second born pulsars and the first born
pulsars that did not accrete any matter. The second, with the lower values of
$\dot{P}$, is filled by the pulsars that have been spun up. This model clearly
struggles to explain the existence of the two observed pulsars with a small
$\dot{P}$,   J1518+4904 and J1829+2456. The second model, APD003T2k, adopts the
values of $\tau_d = 2000{\rm Myrs}$ and $\Delta M_d = 0.0033$ from the best
model of \citet{2008MNRAS.388..393K}.  Note that we do not correlate the initial
spin periods with the magnetic fields.  The lowered $\Delta M_d$ results in all
the pulsars accreting during the CE phase being removed from the visible
population. The top part of the $P-\dot{P}$ diagram corresponds to the pulsars
evolving without any interactions and is very similar to the corresponding
region of the APD05T1k model. The next  model we test is the APD003T2kC model.
It is based on the APD003T2k, but without any accretion in the CE phase allowed.
We do not show a plot for this model as it looks virtually the same as the one
for APD003T2k. This may seem surprising since we argue that the CE phase is the
main process shaping the distribution of pulsars in the $P-\dot{P}$ space for
that model. During RLOF only a small amount of matter is accreted, of the order
of $10^{-5}\;{\rm M_{\sun}}$. This is not enough to significantly affect the
neutron star with a still strong magnetic field.  After the accretion the
star will simply continue its evolution with very similar values of $P$ and
$\dot{P}$, almost as if no accretion occured. Therefore the accreting and the second born pulsar will evolve
essentially likewise. In this model, there would be more pulsars visible in
radio than in the APD003T2k model, but the probability density on the
$P-\dot{P}$ is similar in both cases. The same reasoning applies to the
APD05T1kC modelÊ(that is APD05T1k without any accretion in the CE phase) - it
will look very similar to APD003T1k.

\subsection{Radio-detected fraction}
\label{detfrac}

\begin{figure} \includegraphics{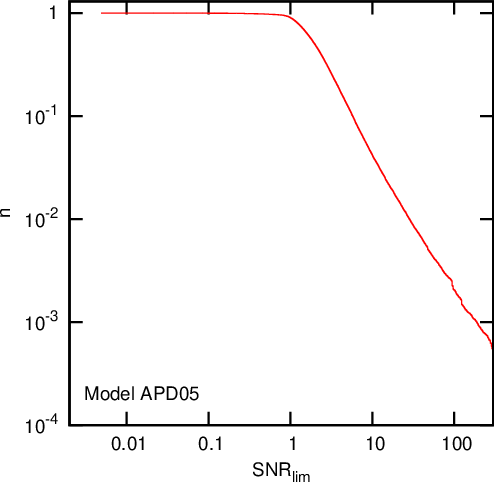} \caption{Cumulative fraction of
pulsars detected as a function of limiting signal-to-noise ratio for the Parkes
Multibeam Pulsar Survey.} \label{detfracplot} \end{figure}

Currently the known sample of double neutron stars is very scarce. Future
instruments like SKA should change this. In Fig. \ref{detfracplot} we present
the fraction of observed radio pulsars as a function of the signal-to-noise
ratio limit for the detection with the Parkes Multibeam Pulsar Survey (PMBPS).
With the realistic limit of 10 sigma, we only see $\approx 5\%$ of the simulated
population. To detect the majority of population we would need to increase the
sensitivity  of detectors by at least a factor of $10$. With a larger  observed
sample  we could check the models much more precisely by repeating the same
analysis as in this work. With the currently available limited sample we are
having problems constraining the evolutionary parameters.  Note that we have
used the parameters of the PMBPS for the selection effects but we simulate
observations of the whole sky. The main for this reason is that choosing a
region of sky that was observed by the PMBPS would significantly reduce our
statistics. Since we neither distinguish between the regions of galaxy when
injecting the simulated pulsars nor correlate the initial spin and/or the
magnetic field with supernova properties, the population observed in any part of
the sky should have equivalent statistical properties as the total population.
We checked that by simulating observations of a smaller patch of sky,
corresponding to the PMBPS, and the results were virtually the same.

\subsection{Comparison with observations}
\label{porow}

In order to  quantify which model of the distribution in the $P-\dot P$ best
describes the data we calculate the likelihood of each model given the data from
Table \ref{pulsobswlas}.  We define the likelihood as :
\begin{equation}
{\cal L}=\prod\limits_{k}^K f(P_k,\dot{P}_k),
\end{equation}
where $P_k$ and $\dot{P}_k$ are the spin period and its derivative values for
the k-th real observed pulsar (see Table \ref{pulsobswlas}) and $f$ is the
probability density defined by Equation \ref{fij} in Section \ref{popnappdot}.
We use two sets of data to calculate the likelihood: the first set excludes both
J0737-3039 A and B as their orbital period is in the range prohibited by one of
our selection effects (see Section \ref{radgw}).  Another special case we
exclude is the B2127+11C pulsar as it probably had a different dynamical history
and does not provide good constraints on the models; the second set does not
contain J0737-3039 B, which is no longer visible, and B2127+11C for the same
reason as above. The latter comparison includes J0737-3039 A as we argue that
the selection effect based on the orbital period is not significant.

\begin{table}
\caption{The logarithm of the likelihood  for all models.  Comparison with
observations: case 1) excluding J0737-3039 A and B and B2127+11C; case 2)
excluding J0737-3039 B and B2127+11C.  The $-\inf$ values originate in
J1518+4904 (APD05T1k) and most pulsars (APD003T2k) lying in the regions were the
modelled probability density is null.}

\label{like}
\begin{tabular}{|c|c|c|}
\hline
 Comparison   &   1 &  2 \\
\hline
A & $-9.82$ & $-11.68$\\
AF & $-12.67$ & $ -16.23 $\\
AP & $-9.72$ & $-11.56$\\
APC & $-10.63$ & $-13.37$\\
APD05 & $-9.59$ & $-11.08$\\
APD05I & $-11.13$ & $-13.49$\\
APD05T1k & $-\inf$ & $-\inf$\\
APD003T2k & $-\inf$ & $-\inf$\\
APD05L & $-11.26$ & $-13.12	$\\
APT20 & $-10.44$ & $ -12.02$\\
HP & $-8.69$ & $-10.87$\\
SP & $-9.74$ & $-11.57$\\
\hline
\end{tabular}
\end{table}

Likelihood values are presented in Table \ref{like}.  In both comparisons with
observations, the model HP and APD05 are the two best models.  The HP model
predicts a very unusual shape of the $P-\dot{P}$ distribution, even though it
has a high likelihood. Another argument against this model is the distribution
of chirp masses, see Section \ref{expected_masses}. Therefore we will focus on
the APD05 as our best model. This is a model with the inefficient propeller
effect taken into account and with increased mass scale for the magnetic field
decay.  The models with the standard value lead to too weak magnetic field of
the synthetic sample. Increasing the magnetic field mass decay scale
$\Delta M_d$ leads to a smaller decay of the field during accretion and
therefore a better agreement with the data. The models with full recycling are
very far from the observations since they predict the population of pulsars in
binaries with spin periods below $20$ms. The models with very long timescales of
the field decay have formally $-\inf$ likelihood as some of the pulsars fall
into bins with no simulated pulsars.  We have verified that the results are not
very sensitive to the number and size of the bins chosen in for the $P- \dot P$
diagram by repeating the likelihood analysis with different values of this
parameters.

We think that this method provides a reasonable quantitative comparison between
models and observations. In our case it is somewhat limited by the small sample
of observed DNSs. For other synthetic population it could prove more useful.

\section{Expected masses of neutron stars observed in  gravitational waves and in the radio}
\label{expected_masses}
\begin{figure*}

\includegraphics[width=0.4\textwidth]{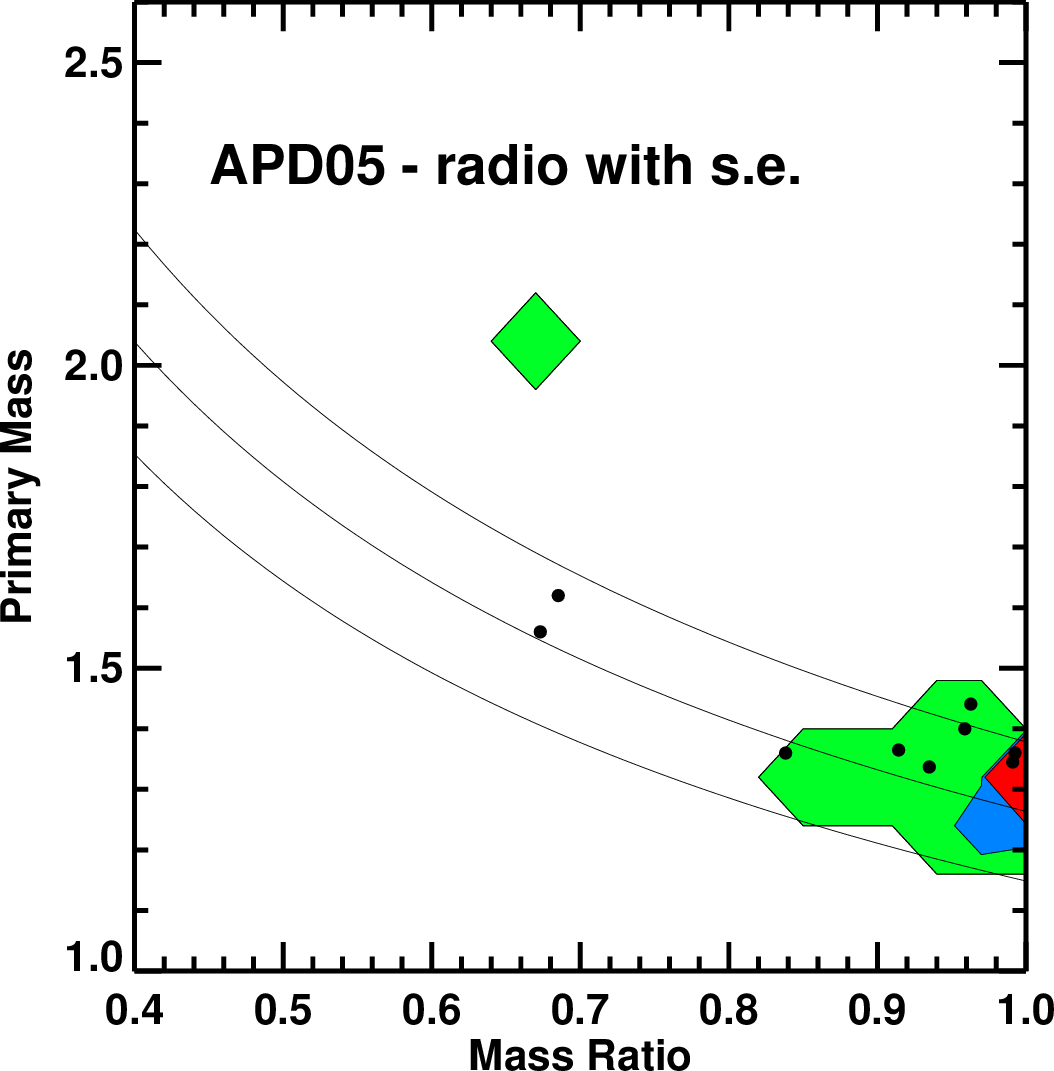}
\includegraphics[width=0.4\textwidth]{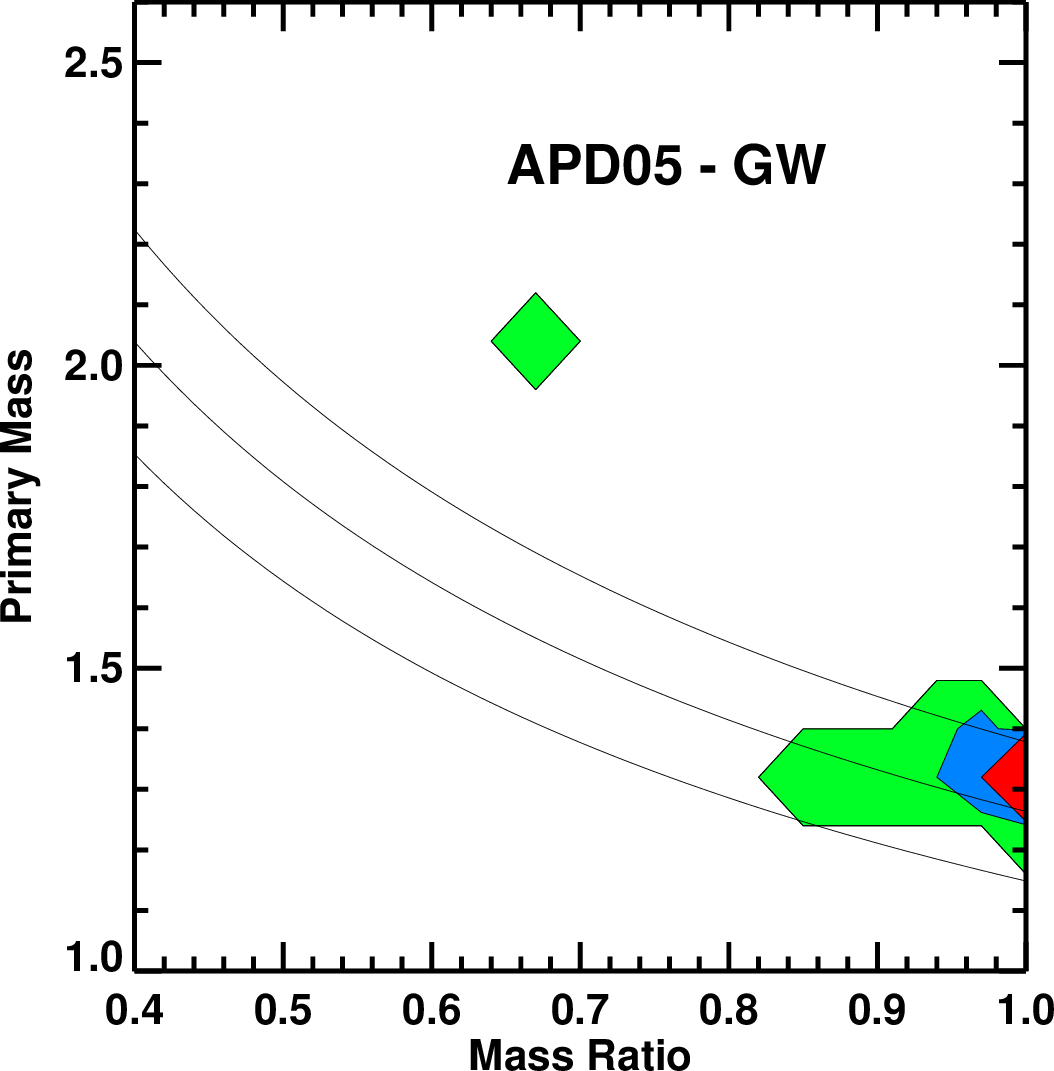}\\
\includegraphics[width=0.4\textwidth]{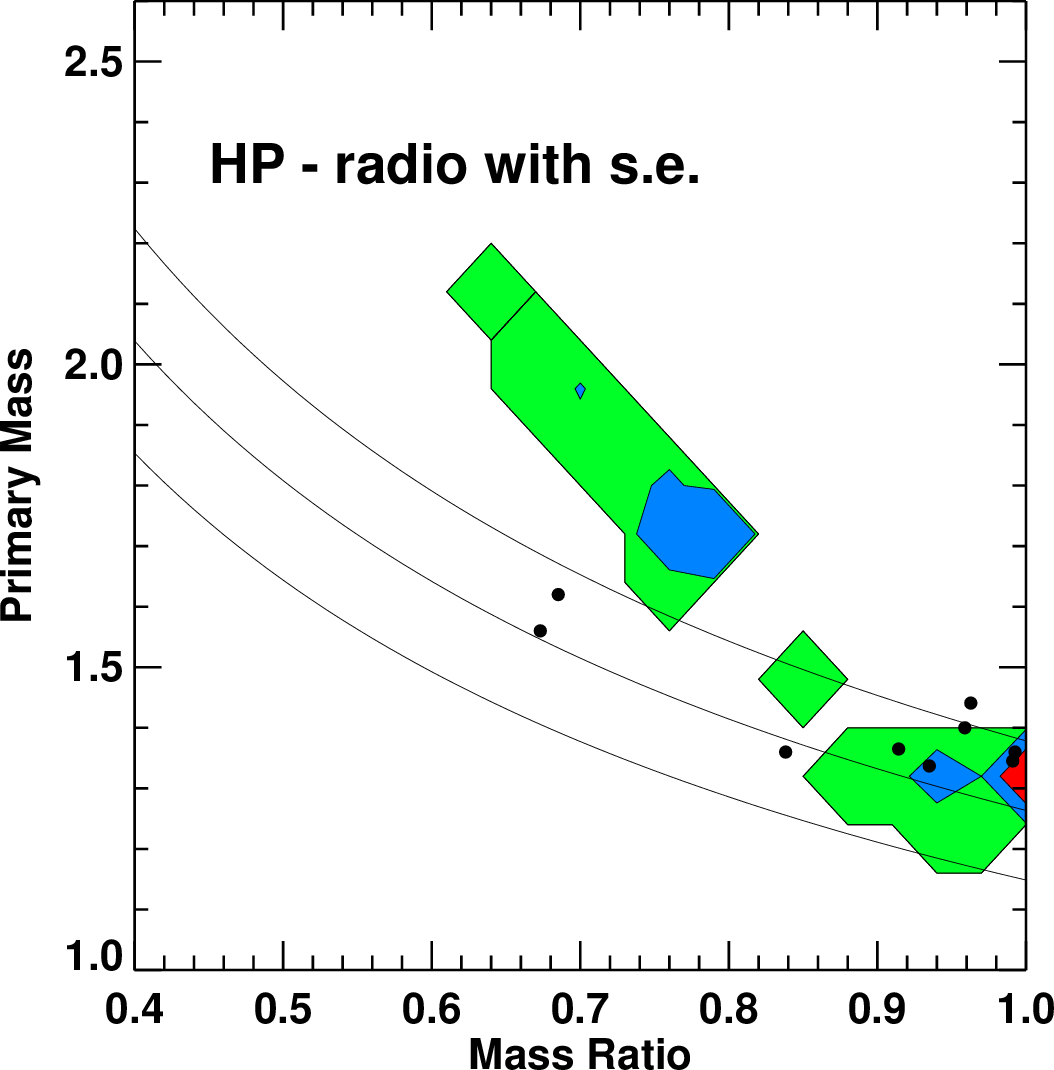}
\includegraphics[width=0.4\textwidth]{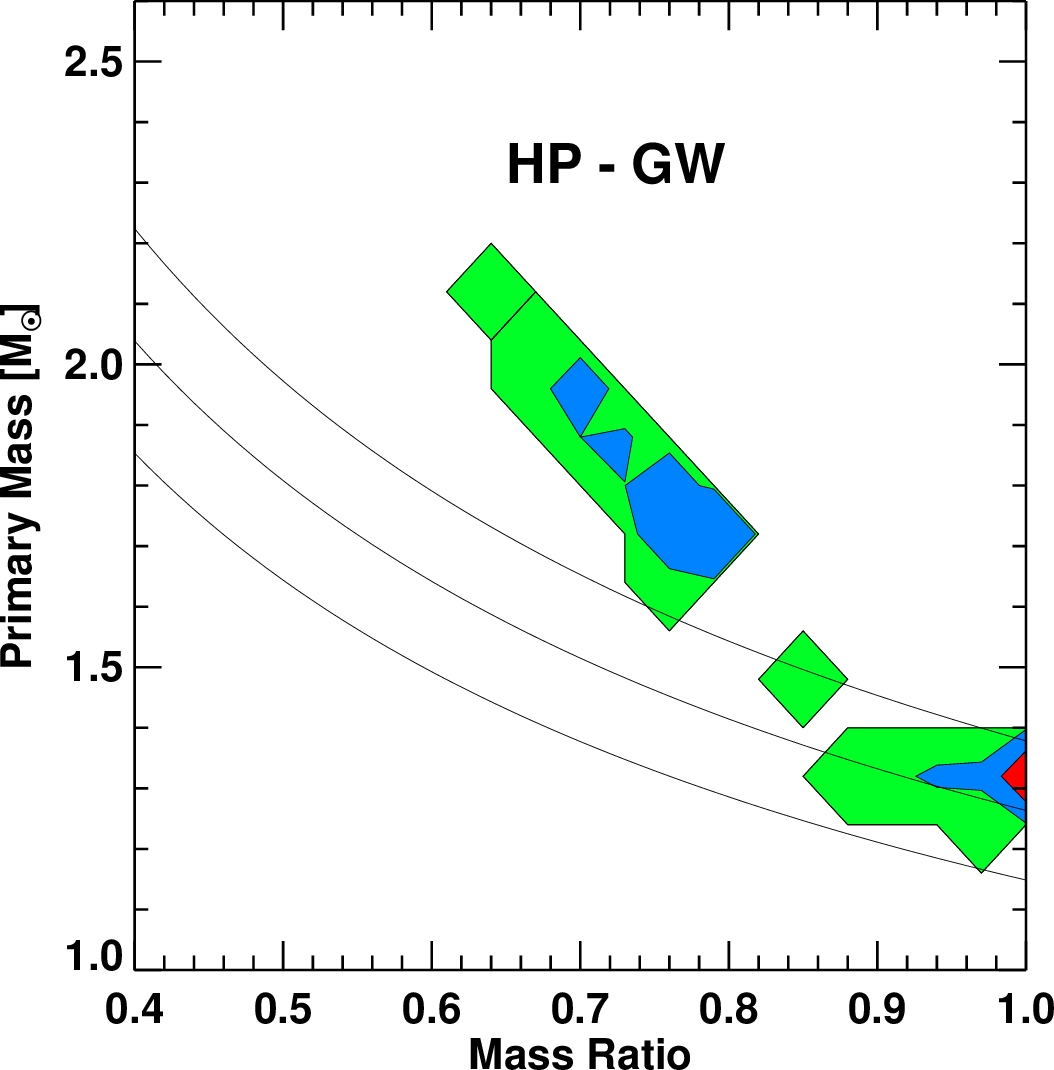}\\
\includegraphics[width=0.4\textwidth]{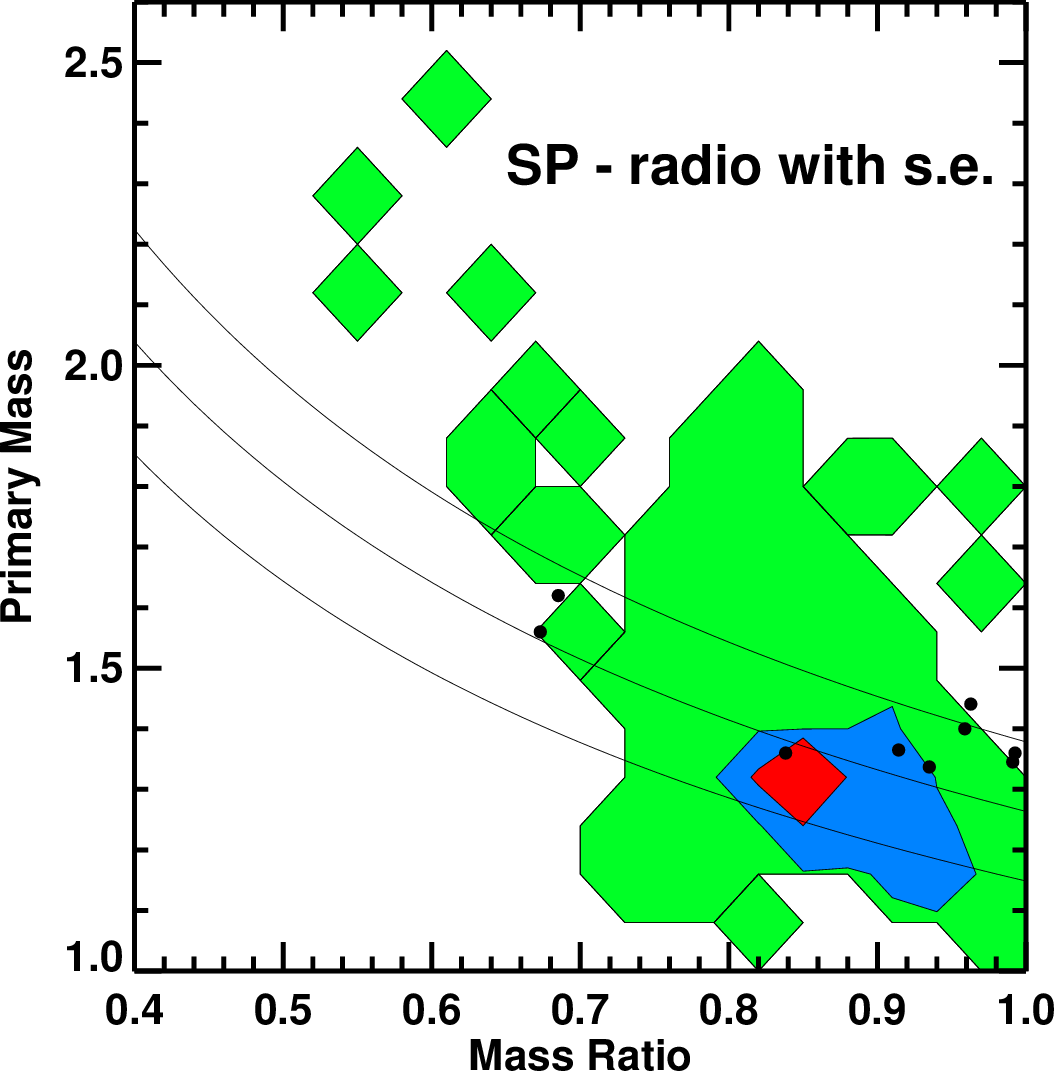}
\includegraphics[width=0.4\textwidth]{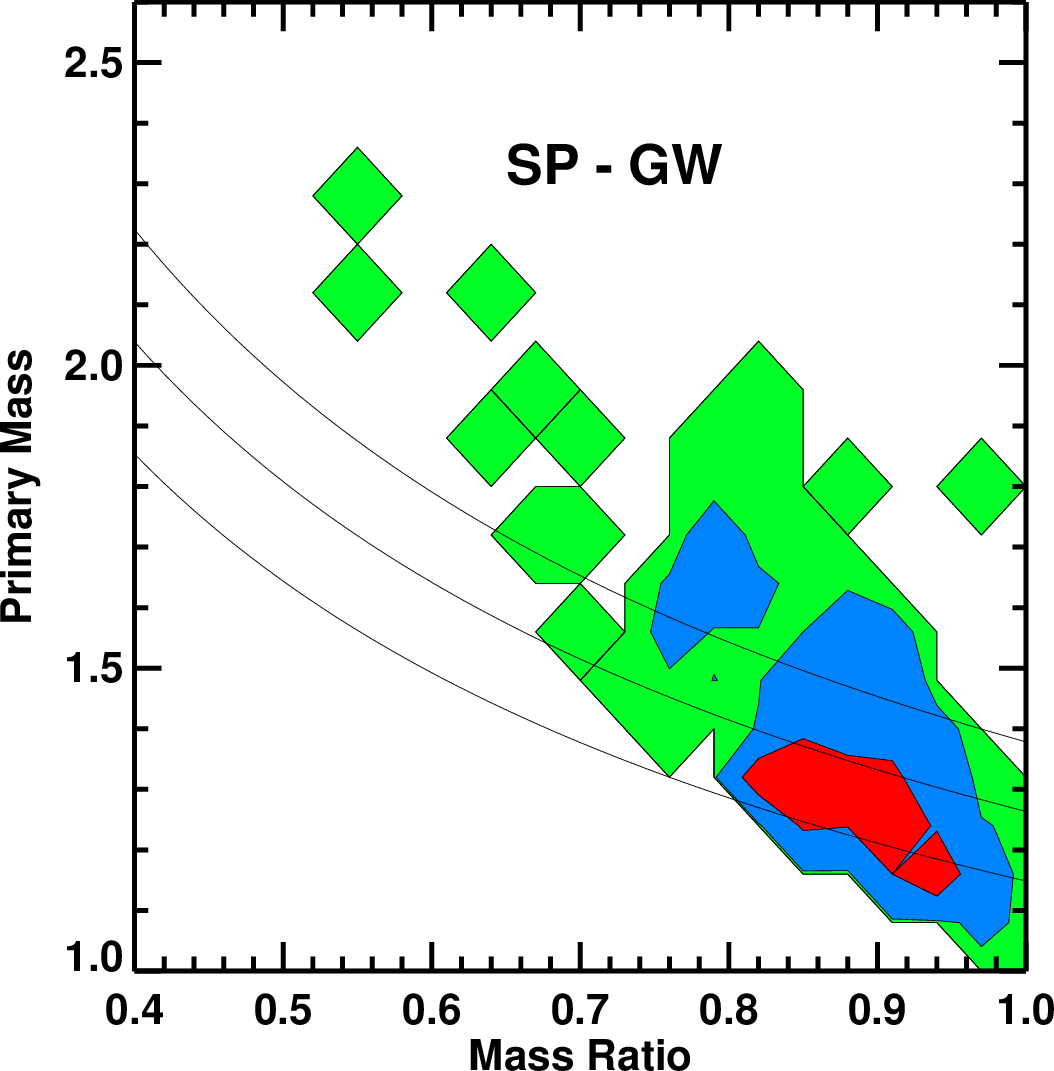}

\caption{Normalised distribution of masses and mass ratios for the simulated
populations observable in the radio (left panels) and gravitational waves (right
panels).  The shadings represents the regions containing 68\%,   95\% , and all
the systems. In each panel we show the lines corresponding to constant chirp
masses of $1.0$, $1.1$, $1.2\,M_{\sun}$. Black points correspond to
observations. }
\label{fig:expmass}

\end{figure*}

In \citet{2004MNRAS.352.1372B} the authors have calculated the expected masses
of neutron stars observed in gravitational waves using the S\textsc{tar}T\textsc{rack}
population synthesis code. The assumed value of the minimum mass and maximum
mass of a neutron star was $1.2 \ M_{\sun}$ and $ 3.0 \ M_{\sun}$ respectively.
They have verified the results by taking the maximum mass to be $2 \ M_{\sun}$
and $2.5 \ M_{\sun}$ \citep{2005MmSAI..76..632G}. The distributions of the mass
ratio $q$ (defined as the ratio of less to more massive component) of neutron
star binaries observed in gravitational waves were shown to have two peaks: the
first one for nearly equal mass systems with both masses close to $ 1.4 \
M_{\sun}$; and the second one with a small mass ratio $q\sim 0.6- 0.7$ in
binaries consisting of a neutron star with gravitational mass close to the
maximum mass with a relatively smaller companion.  The strength of the second
peak depends on the value of the maximum mass of a neutron star.  For the models
with the higher maximal mass the sample starts to be dominated by binaries
belonging to the second peak.  The authors used 20 additional models of stellar
evolution, where they have varied the parameters describing various stages of
stellar and binary evolution in order to assess the robustness of the results.
In that paper, the gravitational wave selected sample was compared with the
volume limited sample. The volume limited sample contained only the binaries
potentially detectable in gravitational waves, i.e. those younger than 10Gyr.

 Up to now there was no detailed	comparison  of properties of neutron
star binaries observed in the radio and gravitational waves.  The only paper
\citep{2005MmSAI..76..513G} that dealt with a similar problem contained a
discussion of radio observability of the gravitational wave selected sample,
within a single model of binary evolution with a wide initial neutron star mass
spectrum.  The authors assumed that the radio selected sample only consists of
binaries with lifetimes (from formation till merger) longer than $100$\,Myrs.
There was no  upper limit on the radio lifetime other than the Hubble time. The
neutron stars in this sample were observable as pulsars for the entire lifetime
of the binary. In that paper it was shown that binary radio pulsars, which
all had long orbital periods, were only a few percent of all binary neutron
stars observed in gravitational waves. The DNSs observed in the radio had a
mass ratio distribution clustered around unity.  The long lived systems  evolved
without the possibility of significant accretion onto a neutron star, while the
short lived systems (with $t_{\rm grav}< 100$\,Myrs) did undergo common envelope
episodes with hypercritical accretion onto the neutron star.  These common
envelope episodes had two consequences: they tightened the orbits and led to a
decrease of the mass ratio of the final double neutron star system, as one of
the neutron stars accreted some matter. This implied that the mass distribution
of the gravitational wave selected sample of double neutron star binaries was
different than the radio selected one.

Here we present a comparison of the radio selected sample of double neutron star
binaries with the observed sample, taking into account  the selection effects.
In this way we choose the best  model and compare the corresponding radio
population with the one selected in gravitational waves.  The radio population
also contains the binaries with merger times greater than the Hubble time that
were omitted in previous efforts. 

In Fig. \ref{fig:expmass} we present the simulated distributions of
objects in the plane spanned by the mass ratio and the primary mass, defined as
the mass of the more massive component of a binary, for APD05, HP and SP
models. The left panels correspond to the radio selected sample, while the
right panels show the gravitational wave selected ones. The objects contained
in the radio sample are weighted by the time they are observable as a pulsar
from Earth, while the objects in the gravitational wave selected sample are
weighted by the volume in which they are detectable (see \ref{gwdet}). The
solid lines correspond to constant values of the chirp mass in these
coordinates. The observed DNS systems are shown as black points. Note that the
masses of neutron stars in the binaries lying in the middle of the plots
(corresponding to J1811-1736 and J1518+4904) are poorly constrained.

\begin{figure}

\includegraphics[width=\columnwidth]{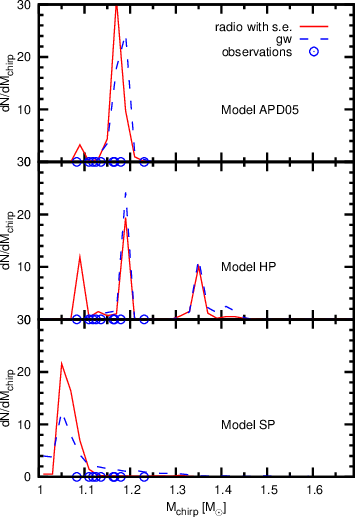}

\caption{The normalised distributions of chirp masses for the simulated
gravitational waves (solid) and radio (dashed) selected samples of double
neutron star binaries. Top panel shows model APD05, middle panel HP and SP is
shown in the bottom panel. Circles on the axes lie where the observed binaries
are. The observed sample is too small to build a distribution.}
\label{fig:ratio}

\end{figure}

In addition, in Fig. \ref{fig:ratio} we  compare the chirp mass distributions of
the binaries selected by their observability in the radio band and  in the
gravitational waves.  Comparing our radio selected sample with the data from
Table~1, we see that the APD05 (top panel) and SP (lower panel) models are more
consistent with the observations than the HP (middle panel) model in which the
majority of binary radio pulsars are predicted to have $q< 0.8$ and one massive
neutron star with $m_2 > 1.6 M_{\sun}$.  The APD05 model, which reproduces best
the observed distribution of pulsars in the $P-\dot P$ space, is consistent with
most of the observed binary radio pulsars. It reproduces the chirp mass
distribution of the radio observed sample well. The low chirp mass systems are
not very likely in this model, but possible. The binary neutron stars with low
mass ratios and moderate masses are not predicted in this model.  However, one
must take into account the fact that the mass measurements of J1811-1736 and
J1518+4904 have quite large error bars.  The SP model is close to reproducing
the distribution of the masses and mass ratios of all observed neutron star
binaries. In this model the distribution of masses is wide. It predicts the
existence of binaries containing massive neutron stars, however the radio
selected distribution is concentrated around $q\sim 0.8-0.9$ and primary masses
$\sim 1.35 M_{\sun}$.

A comparison of the radio selected sample with the gravitational wave one shows
visible differences for SP and HP models and negligible ones for the APD05 model
(see also Fig. \ref{fig:ratio}).  In the APD05 model, the two samples are very
similar, as expected. This is due to the fact that in the standard
S\textsc{tar}T\textsc{rack} model (see Section \ref{astand}) the range of masses of newborn
neutron stars is narrow compared to the previous calculations. In addition, the
amount of matter accreted during the common envelope phase with a helium star is
negligible.

In the HP model both the radio and gravitational wave sample has a large
fraction of  unequal mass binaries with $q=0.7$ and $m_2\approx 1.7 M_{\sun}$.
However, the main peak in the distribution of the chirp mass is for binaries
with both components of about $1.4\;M_{\sun}$.  This model fails to reproduce
the radio observed distribution of masses and mass ratios. At the same time,
this model is the only one able to explain the existence of binaries with a high
chirp mass ($M_{chirp}>1.3 M_{\sun}$), if such binaries exist. It is possible to
form these high mass binaries because of the hypercritical accretion rate. 

In the SP model the gravitational wave selected sample has a tail towards the
higher chirp mass values which is more pronounced with respect to the radio
selected one. This is because of the volume effect:  the sampling volume scales
as $\propto M^{5/2}_{chirp}$.  Thus, the heavier binaries are observable in a
much larger volume in gravitational waves. The mass ratio distribution of the
gravitational wave selected sample leans towards lower values because the
unequal mass binaries typically contain a low mass neutron star with a more
massive companion. They have a higher chirp mass than the equal mass neutron
star binaries, that typically contain two stars with low masses.  This is
similar to the comparison between the gravitational wave selected and the volume
limited sample presented in \citet{2004MNRAS.352.1372B}. The comparison with
radio observations reveals quite an interesting feature: the distribution of
chirp masses is underestimated in this model. However, the observed mass ratios
are relatively well reproduced, as this model leads to a wide distribution of
radio observed mass ratios.

In the APD05 model, the initial masses of the neutron stars come from a very
narrow range.  We do not allow a substantial accretion. Thus most neutron star
binaries have similar masses and there is no possibility to form binaries with
different chirp masses or mass ratios in this model. Hence, there are small differences
between their distributions  for the gravitational wave and radio selected
populations.  This model reproduced the observed chirp masses reasonably well.
The gravitational wave population is shifted towards higher chirp masses due to
the volume effect, as described above for the SP model.

\section{Conclusions}

We have modeled the evolution of binary stars leading to the formation of
double neutron star systems. We then follow their evolution as pulsars and
discuss the properties of the double neutron stars observable in the radio.  We
discuss several models and compare the expected distribution with  observations
in the $P$-$\dot P$ space using likelihood methods that can easily be applied to
other synthesis studies.  In our case, where the observed sample consists of mildly
recycled pulsars, the likelihood method favours different models than when
modelling single pulsars. The second born pulsar evolves identically to solitary ones,
however their contribution to the overall density on the $P$-$\dot{P}$ diagram is small.

First of all, the analysis of the binary star evolution provides little
constraints on the spontaneous magnetic field decay.  This was to be expected as
DNSs are typically observed to be old and evolved objects.  At the same time, we
see that there must be some mechanism at work which leads to a decay of the
magnetic field of the first born star.  This can either be the spontaneous
magnetic field decay, mentioned above, or the magnetic field decay due to
accretion. However, we noticed that in the best fitting models the amount of
accreted matter is small. This is due to the time for the accretion being short,
which is a consequence of the fast evolutionary timescale of the companion.
Moreover, the model with large mass decay scale fits the data better than the
one with a small value.  Thus, in double neutron stars the spontaneous field
decay seems to be much more important than the accretion induced decay.
However, in our model we have assumed that the decay timescale is very short
(few Myr), which does not necessarily provide a good model for the evolution of
isolated pulsars. We argue that the fact that one can get a good fit to the
observations of mildly recycled pulsars in DNS systems suggests that this quick
quenching mimics spin evolution of young pulsars well. Another option present in
the literature is to correlate the initial spin periods with the magnetic field
strength.

The models in which we allow for substantial accretion and recycling of the
neutron star lead to formation of a large number of binaries with periods below
$20$ ms and period derivatives below $10^{-18}\;{\rm s\,s}^{-1}$.  There are no
observed DNSs with such short periods, which means that such substantial
accretion does not take place.  Thus the mass accreted by the first born star
must be small, and the efficiency of recycling during common envelope is very
low.  This efficiency can be much smaller than in the case of a stable mass
transfer.  We do not expect a steady transfer of angular momentum as the mass
transfer can be turbulent.

If the angular momentum transferred is low, then we do not expect double neutron
stars to have pulsation periods  below $\approx 20$ ms.  The binaries containing
pulsars with a shorter period are probably neutron star - white dwarf binaries,
in which the mass transfer had been stable and lasted much longer than in the
progenitors of the double neutron star systems.

 Other studies of synthetic pulsar populations have not made comparisons between
the pulsars detectable by their emission in the radio and these emitting
gravitational waves. We also find that the model resembling the best one from
\citet{2008MNRAS.388..393K} does not produce good results in our case. Double
neutron star systems contain pulsar which are recycled, but only mildly and
would not be called classical recycled pulsars.  It is difficult to produce this
kind of systems in the model based on work by \citet{2008MNRAS.388..393K}.  The
main problem arises from the fact that the pulsar's magnetic field needs to be
quenched without a significant spin-up. Possibly our understanding of the accretion
physics is incomplete and future studies could solve the problem with spinning up
those mildly recycled pulsars.

We compare the properties of the radio selected and gravitational wave selected
sample. We find that the APD05 model is most consistent with observations. It
not only matches the observed sample in the $P-\dot{P}$ space, but it also
reproduces the distribution of chirp masses reasonably well. In this model, the
binary evolution proceeds within the standard S\textsc{tar}T\textsc{rack} model, we allow for
the inefficient propeller effect, partial spin-up proportional to the amount of
mass accreted takes place, and mass scale of magnetic field decay is increased
to $M_d=0.05\,M_{\sun}$.  There are hardly any differences between the
distribution of masses in the radio and gravitational wave selected samples. The
mass ratio of double neutron stars is close to unity in both the radio and
gravitational wave selected sample. However, radio observations show that there
are systems with a mass ratio likely to be in the range of  $0.6 -0.9$, which do
not appear in this model. The presence of such objects can be explained by
assuming the initial mass spectrum of neutron stars is wide, like the one in the
SP model.  The presence of such binaries suggests that the unequal mass systems
with a mass ratio in the range of 0.7 to unity should be included in the search
for gravitational wave signals.  Including a wider initial mass range on one
hand widens the expected distribution of mass ratios, but on the other hand
decreases the agreement between the expected and observed distribution of chirp
masses in the radio. The HP model predicts existence of very high chirp mass
systems, none of which has been discovered. As this model is dominated by the
second born pulsars, this suggests that the observed pulsars are the first born,
mildly recycled objects.

 The models based on the S\textsc{tar}T\textsc{rack} standard model have similar predictions for
the differences between the populations detactable in the radio and
gravitational waves, regardless of the subsequent pulsar evolution details.
Future work should include a bigger observational sample and a wider range of
object types. This should allow to better constrain the pulsar evolution models.
 
 This conclusions may be verified with Square Kilometer Array observations that
should reveal many more DNS systems. The measurement of their masses along with
the identification of the DNS merging population by the Advanced LIGO and VIRGO
will provide further insight into the properties of the population of double
neutron stars.

\section*{Acknowledgments}

This research was supported in part by the Polish Ministry of National Education
(MENiSW) under grants No. N 202 2318 37, 1PO3D00530, N203~302835, N N203 511
238, DPN/N176/VIRGO/200, by the European Gravitational Observatory grant
EGO-DIR-102-2007, by the FOCUS Programme of Foundation for Polish Science and by
CompStar, a Research Networking Programme of the European Science Foundation.
Authors would also like to thank Matthew Bailes for help with improving our
treatment of radio selection effects, sharing the PSREVOLVE code and for his
patience. Discussions about pulsar masses with Ingrid Stairs have been very
fruitful as well. The anonymous referee's comments helped to improve our work
and we found them very useful. We are grateful to Evelyn Caris alias Reynders
and Willem van Straten for their help with the manuscript.

\newcommand{\apj}{ApJ}
\newcommand{\aj}{AJ}
\newcommand{\apjs}{ApJS}
\newcommand{\apjl}{ApJ Lett}
\newcommand{\nat}{Nature}
\newcommand{\aap}{A\&A}
\newcommand{\prc}{Phys.~Rev.~C}
\newcommand{\prd}{Phys.~Rev.~D}
\newcommand{\physrev}{Phys. Rev.}
\newcommand{\mnras}{MNRAS}
\newcommand{\pasp}{PASP}
\newcommand{\pasj}{PASJ}
\newcommand{\apss}{ApSS}
\newcommand{\aapr}{AAPR}
\newcommand{\aaps}{A\&AS}
\newcommand{\physrep}{Phys. Rep.}
\newcommand{\sovast}{Soviet Astron.}
\newcommand{\pasa}{Publ. Astron. Soc. Aust.}
\bibliographystyle{mn2e}
\bibliography{binpuls}

\begin{thebibliography}{}

\bibitem[\protect\citeauthoryear{{Abramovici}, {Althouse}, {Drever}, {Gursel},
  {Kawamura}, {Raab}, {Shoemaker}, {Sievers}, {Spero} \& {Thorne}}{{Abramovici}
  et~al.}{1992}]{1992Sci...256..325A}
{Abramovici} A.,  {Althouse} W.~E.,  {Drever} R.~W.~P.,  {Gursel} Y.,
  {Kawamura} S.,  {Raab} F.~J.,  {Shoemaker} D.,  {Sievers} L.,  {Spero} R.~E.,
     {Thorne} K.~S.,  1992, Science, 256, 325

\bibitem[\protect\citeauthoryear{{Arzoumanian}, {Chernoff} \&
  {Cordes}}{{Arzoumanian} et~al.}{2002}]{2002ApJ...568..289A}
{Arzoumanian} Z.,  {Chernoff} D.~F.,    {Cordes} J.~M.,  2002, \apj, 568, 289

\bibitem[\protect\citeauthoryear{{Arzoumanian}, {Cordes} \&
  {Wasserman}}{{Arzoumanian} et~al.}{1999}]{1999ApJ...520..696A}
{Arzoumanian} Z.,  {Cordes} J.~M.,    {Wasserman} I.,  1999, \apj, 520, 696

\bibitem[\protect\citeauthoryear{{Belczynski} \& {Bulik}}{{Belczynski} \&
  {Bulik}}{2002}]{2002ApJ...574L.147B}
{Belczynski} K.,  {Bulik} T.,  2002, \apjl, 574, L147

\bibitem[\protect\citeauthoryear{{Belczynski}, {Bulik} \&
  {Klu{\'z}niak}}{{Belczynski} et~al.}{2002}]{2002ApJ...567L..63B}
{Belczynski} K.,  {Bulik} T.,    {Klu{\'z}niak} W.~{\l}.,  2002, \apjl, 567,
  L63

\bibitem[\protect\citeauthoryear{{Belczynski}, {Bulik} \& {Rudak}}{{Belczynski}
  et~al.}{2002}]{2002ApJ...571..394B}
{Belczynski} K.,  {Bulik} T.,    {Rudak} B.,  2002, \apj, 571, 394

\bibitem[\protect\citeauthoryear{{Belczy{\'n}ski} \&
  {Kalogera}}{{Belczy{\'n}ski} \& {Kalogera}}{2001}]{2001ApJ...550L.183B}
{Belczy{\'n}ski} K.,  {Kalogera} V.,  2001, \apjl, 550, L183

\bibitem[\protect\citeauthoryear{{Belczynski}, {Kalogera} \&
  {Bulik}}{{Belczynski} et~al.}{2002}]{2002ApJ...572..407B}
{Belczynski} K.,  {Kalogera} V.,    {Bulik} T.,  2002, \apj, 572, 407

\bibitem[\protect\citeauthoryear{{Belczynski}, {Kalogera}, {Rasio}, {Taam},
  {Zezas}, {Bulik}, {Maccarone} \& {Ivanova}}{{Belczynski}
  et~al.}{2008}]{2008ApJS..174..223B}
{Belczynski} K.,  {Kalogera} V.,  {Rasio} F.~A.,  {Taam} R.~E.,  {Zezas} A.,
  {Bulik} T.,  {Maccarone} T.~J.,    {Ivanova} N.,  2008, \apjs, 174, 223

\bibitem[\protect\citeauthoryear{{Benensohn}, {Lamb} \& {Taam}}{{Benensohn}
  et~al.}{1997}]{1997ApJ...478..723B}
{Benensohn} J.~S.,  {Lamb} D.~Q.,    {Taam} R.~E.,  1997, \apj, 478, 723

\bibitem[\protect\citeauthoryear{{Bethe} \& {Brown}}{{Bethe} \&
  {Brown}}{1998}]{1998ApJ...506..780B}
{Bethe} H.~A.,  {Brown} G.~E.,  1998, \apj, 506, 780

\bibitem[\protect\citeauthoryear{{Bhat}, {Cordes}, {Camilo}, {Nice} \&
  {Lorimer}}{{Bhat} et~al.}{2004}]{2004ApJ...605..759B}
{Bhat} N.~D.~R.,  {Cordes} J.~M.,  {Camilo} F.,  {Nice} D.~J.,    {Lorimer}
  D.~R.,  2004, \apj, 605, 759

\bibitem[\protect\citeauthoryear{{Bhattacharya}, {Wijers}, {Hartman} \&
  {Verbunt}}{{Bhattacharya} et~al.}{1992}]{1992A&A...254..198B}
{Bhattacharya} D.,  {Wijers} R.~A.~M.~J.,  {Hartman} J.~W.,    {Verbunt} F.,
  1992, \aap, 254, 198

\bibitem[\protect\citeauthoryear{{Bisnovatyi-Kogan} \&
  {Komberg}}{{Bisnovatyi-Kogan} \& {Komberg}}{1974}]{1974SvA....18..217B}
{Bisnovatyi-Kogan} G.~S.,  {Komberg} B.~V.,  1974, \sovast, 18, 217

\bibitem[\protect\citeauthoryear{{Blaes} \& {Rajagopal}}{{Blaes} \&
  {Rajagopal}}{1991}]{1991ApJ...381..210B}
{Blaes} O.,  {Rajagopal} M.,  1991, \apj, 381, 210

\bibitem[\protect\citeauthoryear{{Bogomazov}, {Abubekerov}, {Lipunov} \&
  {Cherepashchuk}}{{Bogomazov} et~al.}{2005}]{2005ARep...49..295B}
{Bogomazov} A.~I.,  {Abubekerov} M.~K.,  {Lipunov} V.~M.,    {Cherepashchuk}
  A.~M.,  2005, Astronomy Reports, 49, 295

\bibitem[\protect\citeauthoryear{{Bondi} \& {Hoyle}}{{Bondi} \&
  {Hoyle}}{1944}]{1944MNRAS.104..273B}
{Bondi} H.,  {Hoyle} F.,  1944, \mnras, 104, 273

\bibitem[\protect\citeauthoryear{{Bradaschia}, {Calloni}, {Cobal}, {Del
  Fabbro}, {di Virgilio}, {Giazotto}, {Holloway}, {Kautzky}, {Michelozzi},
  {Montelatici}, {Passuello} \& {Velloso}}{{Bradaschia}
  et~al.}{1991}]{1991pfmp.conf..341B}
{Bradaschia} C.,  {Calloni} E.,  {Cobal} M.,  {Del Fabbro} R.,  {di Virgilio}
  A.,  {Giazotto} A.,  {Holloway} L.~E.,  {Kautzky} H.,  {Michelozzi} B.,
  {Montelatici} V.,  {Passuello} D.,    {Velloso} W.,  1991, in Problems of
  Fundamental Modern Physics II {VIRGO: a ground based interferometric antenna
  for gravitational wave detection above 10 Hz.}.
World Scientific, Singapore, pp 341--356

\bibitem[\protect\citeauthoryear{{Brandt} \& {Podsiadlowski}}{{Brandt} \&
  {Podsiadlowski}}{1995}]{1995MNRAS.274..461B}
{Brandt} N.,  {Podsiadlowski} P.,  1995, \mnras, 274, 461

\bibitem[\protect\citeauthoryear{{Brown}, {Lee} \& {Bethe}}{{Brown}
  et~al.}{2000}]{2000ApJ...541..918B}
{Brown} G.~E.,  {Lee} C.-H.,    {Bethe} H.~A.,  2000, \apj, 541, 918

\bibitem[\protect\citeauthoryear{{Bulik} \& {Belczy{\'n}ski}}{{Bulik} \&
  {Belczy{\'n}ski}}{2003}]{2003ApJ...589L..37B}
{Bulik} T.,  {Belczy{\'n}ski} K.,  2003, \apjl, 589, L37

\bibitem[\protect\citeauthoryear{{Bulik}, {Belczy{\'n}ski} \& {Rudak}}{{Bulik}
  et~al.}{2004}]{2004A&A...415..407B}
{Bulik} T.,  {Belczy{\'n}ski} K.,    {Rudak} B.,  2004, \aap, 415, 407

\bibitem[\protect\citeauthoryear{{Bulik}, {Belczy{\'n}ski} \&
  {Zbijewski}}{{Bulik} et~al.}{1999}]{1999MNRAS.309..629B}
{Bulik} T.,  {Belczy{\'n}ski} K.,    {Zbijewski} W.,  1999, \mnras, 309, 629

\bibitem[\protect\citeauthoryear{{Bulik}, {Gondek-Rosinska} \&
  {Belczynski}}{{Bulik} et~al.}{2004}]{2004MNRAS.352.1372B}
{Bulik} T.,  {Gondek-Rosinska} D.,    {Belczynski} K.,  2004, \mnras, 352, 1372

\bibitem[\protect\citeauthoryear{{Buonanno}, {Cook} \& {Pretorius}}{{Buonanno}
  et~al.}{2007}]{2007PhRvD..75l4018B}
{Buonanno} A.,  {Cook} G.~B.,    {Pretorius} F.,  2007, \prd, 75, 124018

\bibitem[\protect\citeauthoryear{{Burgay}, {D'Amico}, {Possenti}, {Manchester},
  {Lyne}, {Joshi}, {McLaughlin}, {Kramer}, {Sarkissian}, {Camilo}, {Kalogera},
  {Kim} \& {Lorimer}}{{Burgay} et~al.}{2003}]{2003Natur.426..531B}
{Burgay} M.,  {D'Amico} N.,  {Possenti} A.,  {Manchester} R.~N.,  {Lyne} A.~G.,
   {Joshi} B.~C.,  {McLaughlin} M.~A.,  {Kramer} M.,  {Sarkissian} J.~M.,
  {Camilo} F.,  {Kalogera} V.,  {Kim} C.,    {Lorimer} D.~R.,  2003, \nat, 426,
  531

\bibitem[\protect\citeauthoryear{{Burgay}, {Joshi}, {D'Amico}, {Possenti},
  {Lyne}, {Manchester}, {McLaughlin}, {Kramer}, {Camilo} \& {Freire}}{{Burgay}
  et~al.}{2006}]{2006MNRAS.368..283B}
{Burgay} M.,  {Joshi} B.~C.,  {D'Amico} N.,  {Possenti} A.,  {Lyne} A.~G.,
  {Manchester} R.~N.,  {McLaughlin} M.~A.,  {Kramer} M.,  {Camilo} F.,
  {Freire} P.~C.~C.,  2006, \mnras, 368, 283

\bibitem[\protect\citeauthoryear{{Choudhuri} \& {Konar}}{{Choudhuri} \&
  {Konar}}{2002}]{2002MNRAS.332..933C}
{Choudhuri} A.~R.,  {Konar} S.,  2002, \mnras, 332, 933

\bibitem[\protect\citeauthoryear{{Contopoulos} \& {Spitkovsky}}{{Contopoulos}
  \& {Spitkovsky}}{2006}]{2006ApJ...643.1139C}
{Contopoulos} I.,  {Spitkovsky} A.,  2006, \apj, 643, 1139

\bibitem[\protect\citeauthoryear{{Cordes} \& {Lazio}}{{Cordes} \&
  {Lazio}}{2002}]{2002astro.ph..7156C}
{Cordes} J.~M.,  {Lazio} T.~J.~W.,  2002, ArXiv Astrophysics e-prints

\bibitem[\protect\citeauthoryear{{Cui}}{{Cui}}{1997}]{1997ApJ...482L.163C}
{Cui} W.,  1997, \apjl, 482, L163+

\bibitem[\protect\citeauthoryear{{Cumming}, {Zweibel} \& {Bildsten}}{{Cumming}
  et~al.}{2001}]{2001ApJ...557..958C}
{Cumming} A.,  {Zweibel} E.,    {Bildsten} L.,  2001, \apj, 557, 958

\bibitem[\protect\citeauthoryear{{D'Angelo} \& {Spruit}}{{D'Angelo} \&
  {Spruit}}{2010}]{2010MNRAS.406.1208D}
{D'Angelo} C.~R.,  {Spruit} H.~C.,  2010, \mnras, 406, 1208

\bibitem[\protect\citeauthoryear{{de Kool}}{{de
  Kool}}{1990}]{1990ApJ...358.189D}
{de Kool} M.,  1990, \apj, 358, 189

\bibitem[\protect\citeauthoryear{{Dewey}, {Taylor}, {Weisberg} \&
  {Stokes}}{{Dewey} et~al.}{1985}]{1985ApJ...294L..25D}
{Dewey} R.~J.,  {Taylor} J.~H.,  {Weisberg} J.~M.,    {Stokes} G.~H.,  1985,
  \apjl, 294, L25

\bibitem[\protect\citeauthoryear{{Dewi}, {Podsiadlowski} \& {Pols}}{{Dewi}
  et~al.}{2005}]{2005MNRAS.363L..71D}
{Dewi} J.~D.~M.,  {Podsiadlowski} P.,    {Pols} O.~R.,  2005, \mnras, 363, L71

\bibitem[\protect\citeauthoryear{{Dewi}, {Podsiadlowski} \& {Sena}}{{Dewi}
  et~al.}{2006}]{2006MNRAS.368.1742D}
{Dewi} J.~D.~M.,  {Podsiadlowski} P.,    {Sena} A.,  2006, \mnras, 368, 1742

\bibitem[\protect\citeauthoryear{{Emmering} \& {Chevalier}}{{Emmering} \&
  {Chevalier}}{1989}]{1989ApJ...345..931E}
{Emmering} R.~T.,  {Chevalier} R.~A.,  1989, \apj, 345, 931

\bibitem[\protect\citeauthoryear{{Faucher-Gigu{\`e}re} \&
  {Kaspi}}{{Faucher-Gigu{\`e}re} \& {Kaspi}}{2006}]{2006ApJ...643..332F}
{Faucher-Gigu{\`e}re} C.-A.,  {Kaspi} V.~M.,  2006, \apj, 643, 332

\bibitem[\protect\citeauthoryear{{Faulkner}}{{Faulkner}}{2004}]{Faulkner}
{Faulkner} A.~J.,  2004, PhD thesis, Faculty of Science and Engineering,
  University of Manchester

\bibitem[\protect\citeauthoryear{{Faulkner}, {Kramer}, {Lyne}, {Manchester},
  {McLaughlin}, {Stairs}, {Hobbs}, {Possenti}, {Lorimer}, {D'Amico}, {Camilo}
  \& {Burgay}}{{Faulkner} et~al.}{2005}]{2005ApJ...618L.119F}
{Faulkner} A.~J.,  {Kramer} M.,  {Lyne} A.~G.,  {Manchester} R.~N.,
  {McLaughlin} M.~A.,  {Stairs} I.~H.,  {Hobbs} G.,  {Possenti} A.,  {Lorimer}
  D.~R.,  {D'Amico} N.,  {Camilo} F.,    {Burgay} M.,  2005, \apjl, 618, L119

\bibitem[\protect\citeauthoryear{Flanagan \& Hughes}{Flanagan \&
  Hughes}{1998}]{Flanagan:1997sx}
Flanagan E.~E.,  Hughes S.~A.,  1998, Phys. Rev., D57, 4535

\bibitem[\protect\citeauthoryear{{Flannery} \& {van den Heuvel}}{{Flannery} \&
  {van den Heuvel}}{1975}]{1975A&A....39...61F}
{Flannery} B.~P.,  {van den Heuvel} E.~P.~J.,  1975, \aap, 39, 61

\bibitem[\protect\citeauthoryear{{Fryer}, {Woosley} \& {Hartmann}}{{Fryer}
  et~al.}{1999}]{1999ApJ...526..152F}
{Fryer} C.~L.,  {Woosley} S.~E.,    {Hartmann} D.~H.,  1999, \apj, 526, 152

\bibitem[\protect\citeauthoryear{{Geppert} \& {Urpin}}{{Geppert} \&
  {Urpin}}{1994}]{1994MNRAS.271..490G}
{Geppert} U.,  {Urpin} V.,  1994, \mnras, 271, 490

\bibitem[\protect\citeauthoryear{{Gnusareva} \& {Lipunov}}{{Gnusareva} \&
  {Lipunov}}{1985}]{1985SvA....29..645G}
{Gnusareva} V.~S.,  {Lipunov} V.~M.,  1985, Soviet Astronomy, 29, 645

\bibitem[\protect\citeauthoryear{{Gondek-Rosi{\'n}ska}, {Bejger}, {Bulik},
  {Gourgoulhon}, {Haensel}, {Limousin}, {Taniguchi} \&
  {Zdunik}}{{Gondek-Rosi{\'n}ska} et~al.}{2007}]{2007AdSpR..39..271G}
{Gondek-Rosi{\'n}ska} D.,  {Bejger} M.,  {Bulik} T.,  {Gourgoulhon} E.,
  {Haensel} P.,  {Limousin} F.,  {Taniguchi} K.,    {Zdunik} L.,  2007,
  Advances in Space Research, 39, 271

\bibitem[\protect\citeauthoryear{{Gondek-Rosi{\'n}ska}, {Bulik} \&
  {Belczy{\'n}ski}}{{Gondek-Rosi{\'n}ska} et~al.}{2005a}]{2005MmSAI..76..632G}
{Gondek-Rosi{\'n}ska} D.,  {Bulik} T.,    {Belczy{\'n}ski} K.,  2005a, Memorie
  della Societa Astronomica Italiana, 76, 632

\bibitem[\protect\citeauthoryear{{Gondek-Rosi{\'n}ska}, {Bulik} \&
  {Belczy{\'n}ski}}{{Gondek-Rosi{\'n}ska} et~al.}{2005b}]{2005MmSAI..76..513G}
{Gondek-Rosi{\'n}ska} D.,  {Bulik} T.,    {Belczy{\'n}ski} K.,  2005b, Memorie
  della Societa Astronomica Italiana, 76, 513

\bibitem[\protect\citeauthoryear{{Gondek-Rosi{\'n}ska}, {Bulik} \&
  {Belczy{\'n}ski}}{{Gondek-Rosi{\'n}ska} et~al.}{2007}]{2007AdSpR..39..285G}
{Gondek-Rosi{\'n}ska} D.,  {Bulik} T.,    {Belczy{\'n}ski} K.,  2007, Advances
  in Space Research, 39, 285

\bibitem[\protect\citeauthoryear{{Gonthier}, {Ouellette}, {Berrier}, {O'Brien}
  \& {Harding}}{{Gonthier} et~al.}{2002}]{2002ApJ...565..482G}
{Gonthier} P.~L.,  {Ouellette} M.~S.,  {Berrier} J.,  {O'Brien} S.,
  {Harding} A.~K.,  2002, \apj, 565, 482

\bibitem[\protect\citeauthoryear{{Gonthier}, {Roberts}, {Nagelkirk}, {Stam} \&
  {Harding}}{{Gonthier} et~al.}{2010}]{2010HEAD...11.1612G}
{Gonthier} P.~L.,  {Roberts} J.~J.,  {Nagelkirk} E.,  {Stam} M.,    {Harding}
  A.~K.,  2010, in Bulletin of the American Astronomical Society Vol.~42 of
  Bulletin of the American Astronomical Society, {Population Synthesis of Radio
  and Gamma-ray Pulsars in the Fermi Era}.
pp 680--+

\bibitem[\protect\citeauthoryear{{Gonthier}, {Van Guilder} \&
  {Harding}}{{Gonthier} et~al.}{2004}]{2004ApJ...604..775G}
{Gonthier} P.~L.,  {Van Guilder} R.,    {Harding} A.~K.,  2004, \apj, 604, 775

\bibitem[\protect\citeauthoryear{{Gunn} \& {Ostriker}}{{Gunn} \&
  {Ostriker}}{1970}]{1970ApJ...160..979G}
{Gunn} J.~E.,  {Ostriker} J.~P.,  1970, \apj, 160, 979

\bibitem[\protect\citeauthoryear{{Hartman}, {Bhattacharya}, {Wijers} \&
  {Verbunt}}{{Hartman} et~al.}{1997}]{1997A&A...322..477H}
{Hartman} J.~W.,  {Bhattacharya} D.,  {Wijers} R.,    {Verbunt} F.,  1997,
  \aap, 322, 477

\bibitem[\protect\citeauthoryear{{Hobbs}, {Lorimer}, {Lyne} \&
  {Kramer}}{{Hobbs} et~al.}{2005}]{2005MNRAS.360..974H}
{Hobbs} G.,  {Lorimer} D.~R.,  {Lyne} A.~G.,    {Kramer} M.,  2005, \mnras,
  360, 974

\bibitem[\protect\citeauthoryear{{Hulse} \& {Taylor}}{{Hulse} \&
  {Taylor}}{1975}]{1975ApJ...195L..51H}
{Hulse} R.~A.,  {Taylor} J.~H.,  1975, \apjl, 195, L51

\bibitem[\protect\citeauthoryear{Hurley, Pols \& Tout}{Hurley
  et~al.}{2000}]{Hurley:2000pk}
Hurley J.~R.,  Pols O.~R.,    Tout C.~A.,  2000, Mon. Not. Roy. Astron. Soc.,
  315, 543

\bibitem[\protect\citeauthoryear{Illarionov \& Sunyaev}{Illarionov \&
  Sunyaev}{1975}]{Illarionov:1975ei}
Illarionov A.~F.,  Sunyaev R.~A.,  1975, Astron. Astrophys., 39, 185

\bibitem[\protect\citeauthoryear{{Jacoby}, {Cameron}, {Jenet}, {Anderson},
  {Murty} \& {Kulkarni}}{{Jacoby} et~al.}{2006}]{2006ApJ...644L.113J}
{Jacoby} B.~A.,  {Cameron} P.~B.,  {Jenet} F.~A.,  {Anderson} S.~B.,  {Murty}
  R.~N.,    {Kulkarni} S.~R.,  2006, \apjl, 644, L113

\bibitem[\protect\citeauthoryear{{Jahan-Miri}}{{Jahan-Miri}}{2000}]{2000ApJ...%
532..514J}
{Jahan-Miri} M.,  2000, \apj, 532, 514

\bibitem[\protect\citeauthoryear{{Janssen}, {Stappers}, {Kramer}, {Nice},
  {Jessner}, {Cognard} \& {Purver}}{{Janssen} et~al.}{2008}]{Janssen}
{Janssen} G.~H.,  {Stappers} B.~W.,  {Kramer} M.,  {Nice} D.~J.,  {Jessner} A.,
   {Cognard} I.,    {Purver} M.~B.,  2008, \aap, 490, 753

\bibitem[\protect\citeauthoryear{{Jorgensen}, {Lipunov}, {Panchenko}, {Postnov}
  \& {Prokhorov}}{{Jorgensen} et~al.}{1995}]{1995Ap&SS.231..389J}
{Jorgensen} H.,  {Lipunov} V.~M.,  {Panchenko} I.~E.,  {Postnov} K.~A.,
  {Prokhorov} M.~E.,  1995, \apss, 231, 389

\bibitem[\protect\citeauthoryear{{Kalogera}, {Narayan}, {Spergel} \&
  {Taylor}}{{Kalogera} et~al.}{2001}]{2001ApJ...556..340K}
{Kalogera} V.,  {Narayan} R.,  {Spergel} D.~N.,    {Taylor} J.~H.,  2001, \apj,
  556, 340

\bibitem[\protect\citeauthoryear{{Kasian}}{{Kasian}}{2008}]{2008AIPC..983..485%
K}
{Kasian} L.,  2008, in {Bassa} C.~G. {Wang}~Z. C.~A.,  M. K.~V.,  eds, 40 Years
  of Pulsars: Millisecond Pulsars, Magnetars and More Vol.~983 of American
  Institute of Physics Conference Series, {Timing and Precession of the Young,
  Relativistic Binary Pulsar PSR J1906+0746}.
pp 485--487

\bibitem[\protect\citeauthoryear{{Kiel} \& {Hurley}}{{Kiel} \&
  {Hurley}}{2009}]{2009MNRAS.395.2326K}
{Kiel} P.~D.,  {Hurley} J.~R.,  2009, \mnras, 395, 2326

\bibitem[\protect\citeauthoryear{{Kiel}, {Hurley}, {Bailes} \& {Murray}}{{Kiel}
  et~al.}{2008}]{2008MNRAS.388..393K}
{Kiel} P.~D.,  {Hurley} J.~R.,  {Bailes} M.,    {Murray} J.~R.,  2008, \mnras,
  388, 393

\bibitem[\protect\citeauthoryear{{Konar} \& {Bhattacharya}}{{Konar} \&
  {Bhattacharya}}{1997}]{1997MNRAS.284..311K}
{Konar} S.,  {Bhattacharya} D.,  1997, \mnras, 284, 311

\bibitem[\protect\citeauthoryear{{Konar} \& {Bhattacharya}}{{Konar} \&
  {Bhattacharya}}{1999a}]{1999MNRAS.303..588K}
{Konar} S.,  {Bhattacharya} D.,  1999a, \mnras, 303, 588

\bibitem[\protect\citeauthoryear{{Konar} \& {Bhattacharya}}{{Konar} \&
  {Bhattacharya}}{1999b}]{1999MNRAS.308..795K}
{Konar} S.,  {Bhattacharya} D.,  1999b, \mnras, 308, 795

\bibitem[\protect\citeauthoryear{{Konar} \& {Choudhuri}}{{Konar} \&
  {Choudhuri}}{2004}]{2004MNRAS.348..661K}
{Konar} S.,  {Choudhuri} A.~R.,  2004, \mnras, 348, 661

\bibitem[\protect\citeauthoryear{{Kuiper}}{{Kuiper}}{1935}]{1935PASP...47...15%
K}
{Kuiper} G.~P.,  1935, \pasp, 47, 15

\bibitem[\protect\citeauthoryear{{Kulczycki}, {Bulik}, {Belczy{\'n}ski} \&
  {Rudak}}{{Kulczycki} et~al.}{2006}]{2006A&A...459.1001K}
{Kulczycki} K.,  {Bulik} T.,  {Belczy{\'n}ski} K.,    {Rudak} B.,  2006, \aap,
  459, 1001

\bibitem[\protect\citeauthoryear{{Large}}{{Large}}{1971}]{1971IAUS...46..165L}
{Large} M.~I.,  1971, in {Davies} R.~D.,  {Graham-Smith} F.,  eds, The Crab
  Nebula Vol.~46 of IAU Symposium, {The Galactic Population of Pulsars}.
Reidel, Dordrecht, pp 165--+

\bibitem[\protect\citeauthoryear{{Lipunov}, {Postnov}, {Prokhorov}, {Panchenko}
  \& {Jorgensen}}{{Lipunov} et~al.}{1995}]{1995ApJ...454..593L}
{Lipunov} V.~M.,  {Postnov} K.~A.,  {Prokhorov} M.~E.,  {Panchenko} I.~E.,
  {Jorgensen} H.~E.,  1995, \apj, 454, 593

\bibitem[\protect\citeauthoryear{{Lorimer}}{{Lorimer}}{2008}]{2008LRR....11...%
.8L}
{Lorimer} D.~R.,  2008, Living Reviews in Relativity, 11, 8

\bibitem[\protect\citeauthoryear{{Lorimer} \& {Kramer}}{{Lorimer} \&
  {Kramer}}{2004}]{2004hpa..book.....L}
{Lorimer} D.~R.,  {Kramer} M.,  2004, {Handbook of Pulsar Astronomy}.
Cambridge University Press

\bibitem[\protect\citeauthoryear{{Lovelace}, {Romanova} \&
  {Bisnovatyi-Kogan}}{{Lovelace} et~al.}{2005}]{2005ApJ...625..957L}
{Lovelace} R.~V.~E.,  {Romanova} M.~M.,    {Bisnovatyi-Kogan} G.~S.,  2005,
  \apj, 625, 957

\bibitem[\protect\citeauthoryear{{Lyne}, {Burgay}, {Kramer}, {Possenti},
  {Manchester}, {Camilo}, {McLaughlin}, {Lorimer}, {D'Amico}, {Joshi},
  {Reynolds} \& {Freire}}{{Lyne} et~al.}{2004}]{2004Sci...303.1153L}
{Lyne} A.~G.,  {Burgay} M.,  {Kramer} M.,  {Possenti} A.,  {Manchester} R.~N.,
  {Camilo} F.,  {McLaughlin} M.~A.,  {Lorimer} D.~R.,  {D'Amico} N.,  {Joshi}
  B.~C.,  {Reynolds} J.,    {Freire} P.~C.~C.,  2004, Science, 303, 1153

\bibitem[\protect\citeauthoryear{{Lyne} \& {Manchester}}{{Lyne} \&
  {Manchester}}{1988}]{1988MNRAS.234..477L}
{Lyne} A.~G.,  {Manchester} R.~N.,  1988, \mnras, 234, 477

\bibitem[\protect\citeauthoryear{{Lyne}, {Manchester} \& {Taylor}}{{Lyne}
  et~al.}{1985}]{1985MNRAS.213..613L}
{Lyne} A.~G.,  {Manchester} R.~N.,    {Taylor} J.~H.,  1985, \mnras, 213, 613

\bibitem[\protect\citeauthoryear{{Lyutikov} \& {Thompson}}{{Lyutikov} \&
  {Thompson}}{2005}]{2005ApJ...634.1223L}
{Lyutikov} M.,  {Thompson} C.,  2005, \apj, 634, 1223

\bibitem[\protect\citeauthoryear{{Manchester}, {Hobbs}, {Teoh} \&
  {Hobbs}}{{Manchester} et~al.}{2005}]{2005yCat.7245....0M}
{Manchester} R.~N.,  {Hobbs} G.~B.,  {Teoh} A.,    {Hobbs} M.,  2005, VizieR
  Online Data Catalog, 7245, 0

\bibitem[\protect\citeauthoryear{{Manchester}, {Lyne}, {Camilo}, {Bell},
  {Kaspi}, {D'Amico}, {McKay}, {Crawford}, {Stairs}, {Possenti}, {Kramer} \&
  {Sheppard}}{{Manchester} et~al.}{2001}]{2001MNRAS.328...17M}
{Manchester} R.~N.,  {Lyne} A.~G.,  {Camilo} F.,  {Bell} J.~F.,  {Kaspi} V.~M.,
   {D'Amico} N.,  {McKay} N.~P.~F.,  {Crawford} F.,  {Stairs} I.~H.,
  {Possenti} A.,  {Kramer} M.,    {Sheppard} D.~C.,  2001, \mnras, 328, 17

\bibitem[\protect\citeauthoryear{{Maron}, {Kijak}, {Kramer} \&
  {Wielebinski}}{{Maron} et~al.}{2000}]{2000A&AS..147..195M}
{Maron} O.,  {Kijak} J.,  {Kramer} M.,    {Wielebinski} R.,  2000, \aaps, 147,
  195

\bibitem[\protect\citeauthoryear{{Melatos} \& {Phinney}}{{Melatos} \&
  {Phinney}}{2001}]{2001PASA...18..421M}
{Melatos} A.,  {Phinney} E.~S.,  2001, \pasa, 18, 421

\bibitem[\protect\citeauthoryear{{Miyamoto} \& {Nagai}}{{Miyamoto} \&
  {Nagai}}{1975}]{1975PASJ...27..533M}
{Miyamoto} M.,  {Nagai} R.,  1975, \pasj, 27, 533

\bibitem[\protect\citeauthoryear{{Muslimov} \& {Tsygan}}{{Muslimov} \&
  {Tsygan}}{1985}]{1985Ap&SS.115...43M}
{Muslimov} A.~G.,  {Tsygan} A.~I.,  1985, \apss, 115, 43

\bibitem[\protect\citeauthoryear{{Narayan} \& {Ostriker}}{{Narayan} \&
  {Ostriker}}{1990}]{1990ApJ...352..222N}
{Narayan} R.,  {Ostriker} J.~P.,  1990, \apj, 352, 222

\bibitem[\protect\citeauthoryear{{Ostriker} \& {Gunn}}{{Ostriker} \&
  {Gunn}}{1969}]{1969ApJ...157.1395O}
{Ostriker} J.~P.,  {Gunn} J.~E.,  1969, \apj, 157, 1395

\bibitem[\protect\citeauthoryear{{Paczynski}}{{Paczynski}}{1990}]{1990ApJ...34%
8..485P}
{Paczynski} B.,  1990, \apj, 348, 485

\bibitem[\protect\citeauthoryear{{Perera}, {McLaughlin}, {Kramer}, {Stairs},
  {Ferdman}, {Freire}, {Possenti}, {Breton}, {Manchester}, {Burgay}, {Lyne} \&
  {Camilo}}{{Perera} et~al.}{2010}]{2010ApJ...721.1193P}
{Perera} B.~B.~P.,  {McLaughlin} M.~A.,  {Kramer} M.,  {Stairs} I.~H.,
  {Ferdman} R.~D.,  {Freire} P.~C.~C.,  {Possenti} A.,  {Breton} R.~P.,
  {Manchester} R.~N.,  {Burgay} M.,  {Lyne} A.~G.,    {Camilo} F.,  2010, \apj,
  721, 1193

\bibitem[\protect\citeauthoryear{Peters}{Peters}{1964}]{PhysRev.136.B1224}
Peters P.~C.,  1964, Phys. Rev., 136, B1224

\bibitem[\protect\citeauthoryear{{Phinney} \& {Sigurdsson}}{{Phinney} \&
  {Sigurdsson}}{1991}]{1991Natur.349..220P}
{Phinney} E.~S.,  {Sigurdsson} S.,  1991, \nat, 349, 220

\bibitem[\protect\citeauthoryear{{Piran}}{{Piran}}{1992}]{1992AIPC..272.1626P}
{Piran} T.,  1992, in R. S.~J.,  ed., American Institute of Physics Conference
  Series Vol.~272 of American Institute of Physics Conference Series,
  {{$\gamma$}-ray bursts and neutron star mergers-possibly the strongest
  explosions in the universe}.
pp 1626--1633

\bibitem[\protect\citeauthoryear{{Popov}, {Colpi}, {Treves}, {Turolla},
  {Lipunov} \& {Prokhorov}}{{Popov} et~al.}{2000}]{2000A&AT...19..471P}
{Popov} S.~B.,  {Colpi} M.,  {Treves} A.,  {Turolla} R.,  {Lipunov} V.~M.,
  {Prokhorov} M.~E.,  2000, Astronomical and Astrophysical Transactions, 19,
  471

\bibitem[\protect\citeauthoryear{{Portegies Zwart} \& {Yungelson}}{{Portegies
  Zwart} \& {Yungelson}}{1998}]{1998A&A...332..173P}
{Portegies Zwart} S.~F.,  {Yungelson} L.~R.,  1998, \aap, 332, 173

\bibitem[\protect\citeauthoryear{{Quinn}, {Perrine}, {Richardson} \&
  {Barnes}}{{Quinn} et~al.}{2010}]{2010AJ....139..803Q}
{Quinn} T.,  {Perrine} R.~P.,  {Richardson} D.~C.,    {Barnes} R.,  2010, \aj,
  139, 803

\bibitem[\protect\citeauthoryear{{Ransom}, {Cordes} \& {Eikenberry}}{{Ransom}
  et~al.}{2003}]{2003ApJ...589..911R}
{Ransom} S.~M.,  {Cordes} J.~M.,    {Eikenberry} S.~S.,  2003, \apj, 589, 911

\bibitem[\protect\citeauthoryear{{Romani}}{{Romani}}{1990}]{1990Natur.347..741%
R}
{Romani} R.~W.,  1990, \nat, 347, 741

\bibitem[\protect\citeauthoryear{{Romanova}, {Ustyugova}, {Koldoba} \&
  {Lovelace}}{{Romanova} et~al.}{2004}]{2004ApJ...616L.151R}
{Romanova} M.~M.,  {Ustyugova} G.~V.,  {Koldoba} A.~V.,    {Lovelace} R.~V.~E.,
   2004, \apjl, 616, L151

\bibitem[\protect\citeauthoryear{{Rudak} \& {Ritter}}{{Rudak} \&
  {Ritter}}{1994}]{1994MNRAS.267..513R}
{Rudak} B.,  {Ritter} H.,  1994, \mnras, 267, 513

\bibitem[\protect\citeauthoryear{{Ruderman}}{{Ruderman}}{1991a}]{1991ApJ...366%
..261R}
{Ruderman} M.,  1991a, \apj, 366, 261

\bibitem[\protect\citeauthoryear{{Ruderman}}{{Ruderman}}{1991b}]{1991ApJ...382%
..587R}
{Ruderman} M.,  1991b, \apj, 382, 587

\bibitem[\protect\citeauthoryear{{Ruderman}}{{Ruderman}}{1991c}]{1991ApJ...382%
..576R}
{Ruderman} R.,  1991c, \apj, 382, 576

\bibitem[\protect\citeauthoryear{{Scalo}}{{Scalo}}{1986}]{1986FCPh...11....1S}
{Scalo} J.~M.,  1986, Fundamentals of Cosmic Physics, 11, 1

\bibitem[\protect\citeauthoryear{{Shibazaki}, {Murakami}, {Shaham} \&
  {Nomoto}}{{Shibazaki} et~al.}{1989}]{1989Natur.342..656S}
{Shibazaki} N.,  {Murakami} T.,  {Shaham} J.,    {Nomoto} K.,  1989, \nat, 342,
  656

\bibitem[\protect\citeauthoryear{{Stairs}}{{Stairs}}{2004}]{2004Sci...304..547%
S}
{Stairs} I.~H.,  2004, Science, 304, 547

\bibitem[\protect\citeauthoryear{{Stollman}}{{Stollman}}{1987}]{1987A&A...178.%
.143S}
{Stollman} G.~M.,  1987, \aap, 178, 143

\bibitem[\protect\citeauthoryear{{Story}, {Gonthier} \& {Harding}}{{Story}
  et~al.}{2007}]{2007ApJ...671..713S}
{Story} S.~A.,  {Gonthier} P.~L.,    {Harding} A.~K.,  2007, \apj, 671, 713

\bibitem[\protect\citeauthoryear{{Taam} \& {van den Heuvel}}{{Taam} \& {van den
  Heuvel}}{1986}]{1986ApJ...305..235T}
{Taam} R.~E.,  {van den Heuvel} E.~P.~J.,  1986, \apj, 305, 235

\bibitem[\protect\citeauthoryear{{Tauris} \& {Manchester}}{{Tauris} \&
  {Manchester}}{1998}]{1998MNRAS.298..625T}
{Tauris} T.~M.,  {Manchester} R.~N.,  1998, \mnras, 298, 625

\bibitem[\protect\citeauthoryear{{Taylor} \& {Manchester}}{{Taylor} \&
  {Manchester}}{1977}]{1977ApJ...215..885T}
{Taylor} J.~H.,  {Manchester} R.~N.,  1977, \apj, 215, 885

\bibitem[\protect\citeauthoryear{{Timmes}, {Woosley} \& {Weaver}}{{Timmes}
  et~al.}{1996}]{1996ApJ...457..834T}
{Timmes} F.~X.,  {Woosley} S.~E.,    {Weaver} T.~A.,  1996, \apj, 457, 834

\bibitem[\protect\citeauthoryear{{Urpin} \& {Geppert}}{{Urpin} \&
  {Geppert}}{1995}]{1995MNRAS.275.1117U}
{Urpin} V.,  {Geppert} U.,  1995, \mnras, 275, 1117

\bibitem[\protect\citeauthoryear{{Urpin}, {Konenkov} \& {Urpin}}{{Urpin}
  et~al.}{1997}]{1997MNRAS.292..167U}
{Urpin} V.,  {Konenkov} D.,    {Urpin} V.,  1997, \mnras, 292, 167

\bibitem[\protect\citeauthoryear{{Vanbeveren}, {De Loore} \& {Van
  Rensbergen}}{{Vanbeveren} et~al.}{1998}]{1998A&ARv...9...63V}
{Vanbeveren} D.,  {De Loore} C.,    {Van Rensbergen} W.,  1998, \aapr, 9, 63

\bibitem[\protect\citeauthoryear{{Weaver} \& {Woosley}}{{Weaver} \&
  {Woosley}}{1993}]{1993PhR...227...65W}
{Weaver} T.~A.,  {Woosley} S.~E.,  1993, \physrep, 227, 65

\bibitem[\protect\citeauthoryear{{Webbink}}{{Webbink}}{1984}]{1984ApJ...277..3%
55W}
{Webbink} R.~F.,  1984, \apj, 277, 355

\bibitem[\protect\citeauthoryear{{Weisberg} \& {Taylor}}{{Weisberg} \&
  {Taylor}}{2005}]{2005ASPC..328...25W}
{Weisberg} J.~M.,  {Taylor} J.~H.,  2005, in {Rasio} F.~A.,  {Stairs} I.~H.,
  eds, Binary Radio Pulsars Vol.~328 of Astronomical Society of the Pacific
  Conference Series, {The Relativistic Binary Pulsar B1913+16: Thirty Years of
  Observations and Analysis}.
pp 25--+

\bibitem[\protect\citeauthoryear{{Woosley}}{{Woosley}}{1986}]{1986nce..conf...%
.1W}
{Woosley} S.~E.,  1986, in {Audouze} J.,  {Chiosi} C.,   {Woosley} S.~E.,  eds,
  Saas-Fee Advanced Course 16: Nucleosynthesis and Chemical Evolution
  {Nucleosynthesis and Stellar Evolution}.
Geneva Observatory, CH-190 Sauverny, pp~1--+

\bibitem[\protect\citeauthoryear{{Young}, {Manchester} \& {Johnston}}{{Young}
  et~al.}{1999}]{1999Natur.400..848Y}
{Young} M.~D.,  {Manchester} R.~N.,    {Johnston} S.,  1999, \nat, 400, 848

\bibitem[\protect\citeauthoryear{{Zhang} \& {Kojima}}{{Zhang} \&
  {Kojima}}{2006}]{2006MNRAS.366..137Z}
{Zhang} C.~M.,  {Kojima} Y.,  2006, \mnras, 366, 137

\end{thebibliography}

\label{lastpage}

\end{document}